\documentclass[journal]{IEEEtran}
\usepackage{graphicx}
\usepackage{color}

\ifCLASSINFOpdf
\else
\fi
\usepackage[utf8]{inputenc}
\usepackage{amsmath}
\usepackage[numbers]{natbib}
\usepackage{rotating}
\usepackage{adjustbox}
\usepackage[graphicx]{realboxes}
\usepackage{longtable}
\usepackage{natbib,hyperref}
\usepackage{amssymb}
\begin{document}
%
\title{Prototype Filters and Filtered Waveforms for
Radio Air-Interfaces: A Review}
%
%
%

\author{M.Y BENDIMERAD*, S.SENHADJI and F.T.BENDIMERAD \\ *Department of Electrical Engineering, University of Bechar \\ LTT Laboratory, University of Tlemcen}
\maketitle


\begin{abstract}
The perspective for the next generation of wireless communications prognosticates that the communication should take place with high heterogeneity in terms of services and specifications. The emergence of new communication networks, especially between devices like internet of things (IoT) networks, machine type communications (MTC) and unmanned aerial vehicles (UAV), makes the redesign of the radio air-interfaces even higher significant. High data rate, efficient use of the spectrum, low latency, flexibility, scalable ability, energy efficiency and reduced complexity are in the vanguard of the challenging needs. This highlights the necessity to overrides classical waveforms and propose new air interface that best fits these requirements. Several attempts were conducted to propose the most suitable prototype filters and their associated waveforms that support the future applications with these crucial prerequisites. In the present paper, we analyze the most important marked prototype filters reported in literature, and their corresponding waveforms proposed as prominent candidates for wireless air interface. We will show that the filter properties impact the waveform performances and that a close filter-waveform interrelationship is there. The principal contribution is to give a deep insight on the structure of these filters and waveforms and to come up with a fair comparison regarding different criteria and distinct figures of merit.
\end{abstract}


%
\IEEEpeerreviewmaketitle

\section{Introduction}

Since the first generation, mobile wireless communication systems  have known  a steady evolution. Four generations of the cellular networks have been established and standardized up to now, initiated by the first analogue generation where only voice communication was ensured (AMPS), they evolved then toward digital generations: Global System for Mobile Communications (GSM), General Packet Radio Service (GPRS), Enhanced Data Rates for GSM Evolution (EDGE), Universal Mobile Telecommunications System (UMTS), Long-Term Evolution (LTE) and LTE-A, all characterized by high capacity, diversity and quality of services. The digital processing of the data has contributed significantly  to this evolution in the sense that it makes information more flexible to be handled. The emergence of new applications,  like internet of things (IoT), e-healthcare, device to device (D2D), machine to machine (M2M), unmanned aerial vehicles (UAV) and machine type communications (MTC) are creating a significant challenge on cellular networks. These are expected to support of tens of thousands of devices connected simultaneously {\cite{ghavimi2014m2m}\cite{perera2013context}. For instance, the ultra-reliable low latency communications (URLLC) requires high reliability, a data rate about $1$Gpbs, an ultra-accurate  device positioning and latency less than $1$ ms. The enhanced mobile broadband (eMBB) scenario should maintain very large bandwidth allocation in the range of $6-90$ GHz with a data rate up to $10$ Gbps. Also, D2D applications are expected to enable a direct communication between devices in proximity, they require therefore a high energy efficiency and a low latency. Hence, the primary concern is to satisfy the exponential rise in user and traffic capacity. On the other hand, it is essential to decrease the latency and the control signaling to ensure a great autonomy of the device \cite{asadi2014survey}. To this concern, new research areas are explored, like Massive Multiple-Input Multiple-Output (mMIMO), small cells, Coordinated Multi-Point (CoMP), millimeter wave (mmwave) communications and beam-forming techniques.  These approaches are essential for 5G and beyond 5G (B5G) wireless networks. B5G technology is expected to be a leading infrastructure for communication industry with a focus on intelligence involvement. Artificial intelligence (AI) and machine learning (ML) are increasingly incorporated in solving problems dealing with the physical layer interface. More specifically, ML and AI were recently used in channel learning and path-loss prediction, as well as in latency and energy consumption reduction \cite{morocho2019machine}. 

To support all these technologies, different physical-layer air interfaces have been investigated and proposed in literature. The air interface specifies procedures of adapting information to a wireless communication medium. The symbols transmitted via an air-interface belong to a constellation, i.e., Phase-shift keying (PSK) or Quadrature amplitude modulation (QAM). In this case, shaping pulses must maintain maximum spectral efficiency. A pulse shape or prototype filter describes how energy is distributed (in time, frequency)  or any other dimension. It is a key factor to show the energy dispersion of the signal and to identify the multi-carrier structure, i.e., orthogonal, non-orthogonal or bi-orthogonal. If the prototype filter involved at the transmitted side and the received side, do not show any correlation in all time-frequency (TF) lattice points, the scheme is either bi-orthogonal or orthogonal. When orthogonal structures maximize the signal to noise (SNR), bi-orthogonal structures may reach better performance for dispersive channels. In literature, different criteria are adopted in the conception of prototype filters: energy concentration (\cite{strohmer2003optimal},\cite{landau1961prolate}), transition band decay  \cite{martin1998small}, \cite{haas1997time} and hardware implementation \cite{guan2020bidirectional}, etc. The rectangular pulse, although perfectly localized in time, shows a bad localization in frequency. This produces interference between carriers in frequency selective channels \cite{strohmer2003optimal}. To overcome this drawback, it is possible to lengthen the time symbol, as for sub-critical density or over-sampled modulations \cite{vallet1995fraction}. In this case, a decrease in spectral efficiency is noted. Using short-time pulse shape, like rectangular function, reduces the computational complexity and communications latency. Nevertheless, these finite length filters cause high spectral leakage. To cope with, prolate spheroidal wave functions (PSWFs) were proposed. They are time-limited filters with small side-lobes \cite{walter2005new}, \cite{moore2004prolate}. The Hermite$-$Gaussian functions   are special cases of PSWFs which provide optimum TF concentration. However, these functions don't satisfy the Nyquist criterion. A linear deformation of the Gaussian pulse by Hermit functions leads to the design of Hermite filter \cite{haas1997time}. By this way, the Nyquist criterion is fulfilled.  Half Square Root Raised Cosine Function (SRRC) is devoted to satisfy Nyquist criterion after matching  filtering. It takes advantage from the excellent compromise of TF behavior \cite{farhang2010cosine}, \cite{siohan2002analysis}. The Isotropic Orthogonal Transform Algorithm (IOTA) keeps the good concentration property of Gaussian filter, and adds orthogonality to prevent adjacent interference \cite{le1995coded} . It is shown in \cite{duffin1952class} that an analytical expression can be derived for the IOTA algorithm, which makes the function more tractable. This pulse is called extended Gaussian function (EGF). It achieves exactly the same responses in time and frequency domain \cite{roche1997family}. Mirabbasi-Martin functions were proposed to ensure a good transition band decaying property. They exhibit the distinctive that their derivative functions are continuous \cite{mirabbasi2002design}.The European Physical layer for Dynamic Access (PHYDYAS) project investigates on the possibility of adapting Mirabbasi-Martin functions to multi-carrier filter bank scheme. 

Multi-carrier modulations have the advantage to be robust to multi-path channels. Orthogonal frequency division multiplexing (OFDM) consists in the division of the overall transmission bandwidth into several narrower parallel sub-channels, each of them transmitting a lower-rate signal, which can be decoded efficiently through a simple one-tap equalizer \cite{payaro2011resource} \cite{nee2000ofdm}.  Although OFDM technique has been adopted as the air interface in several wireless communication standards, it exhibits some intrinsic drawbacks. These make OFDM unsuitable to be adopted in specific contexts as double dispersive channel environments \cite{farhang2011ofdm}. The spectral efficiency (SE) loss caused by cyclic prefix insertion and the synchronization required for orthogonality are the major shortcomings for the orthogonal OFDM signaling technique. In these cases, the low complexity of OFDM technique would wear off as synchronization algorithms are inducted with high-level of calculation \cite{nugraha2016performance}. OFDM also suffers from other critical problems such as the great peak to average power ratio (PAPR) when the number of sub-carriers is important and the high side lobes due to the rectangular transmitting filter. Some solutions have been proposed to overcome the aforementioned OFDM's drawbacks, like relaxing the tight synchronization by proposing algorithms for multi-user interference cancellation \cite{lee2012mmse} \cite{lee2011cfo}, or for reducing out-of-band leakage \cite{huang2015out}. However, all these methods are generally very complex to implement and they increase the receiver calculations. In these circumstances, Filtered Bank Multi Carrier (FBMC) modulation was proposed to replace OFDM \cite{siohan2002analysis}. While relying on dividing the spectrum into multiple sub-bands, too, FBMC applies a filtering function to each of the sub-carriers. Therefore, the side-lobes are much weaker and thus the inter-carrier interference (ICI) issue is by far less crucial than with OFDM \cite{schaich2014waveform}. Therefore, a system applying FBMC is much more suited to a potential 5G or beyond 5G (B5G) system in that respect \cite{schaich2014waveform} \cite{bellanger2010fbmc}. Although FBMC theory reveals a set of valorous features, most of them are practically neglected, due their no feasibility and difficult implementation. Recent tendencies try to design waveforms or their equivalent prototype filters that reunite the theoretical advantages and a simple practical implementation\cite{viholainen2009prototype}  \cite{bellanger2001specification} \cite{tabatabaee2017prototype} \cite{zhang2016polyphase}. New waveforms using well-shaped filters have achieved great attention recently. These are : generalized frequency division multiplexing (GFDM) \cite{michailow2014generalized}, universal filtered multi-carrier (UFMC) \cite{vakilian2013universal}[11] filtered orthogonal frequency division multiplexing (F-OFDM) \cite{xiao2010transmit} and windowed orthogonal frequency division multiplexing (W-OFDM) \cite{bala2013shaping}.

In the subsequent paragraphs, we will try to analyze and compare the performances of all these prototype filers and all these multi-carrier modulation schemes. The major contributions of the present paper are listed as follows: 

\begin{list}{$-$}{\topsep=6pt \itemsep=0pt \leftmargin=10pt}
\item A review of the state of the art regarding the mathematical formulations and the design methods of filter functions.

\item An investigation on the selection criteria between prototype filters. 

\item A review of the most prominent filtered multi-carrier modulation (MCM) schemes meeting the aforementioned requirements.

\item An exhaustive comparison between MCM schemes based on several key performance indicators and within different scenarios. 
 \end{list}

The paper is outline as follows: section (2) reviews the state-of-the art of prototype filters. Section (3) assesses the filters performance and the selection criteria. Section (4) approaches the different filtered waveforms from a structural point of view. In section (5), we discuss  the assessment of these waveforms in different contexts and under various conditions. Meaningful outcomes on the filter-waveform dependency in respect of this review are marked in section (6). Section (7) concludes this survey.

\paragraph{Notations}
For best clearness and readability, we define in this part the notations used in the subsequent sections. These have for meaning that specified in the following table; except if in the text it is mentioned that the notation has another connotation.

\begin{center}
\begin{table}[h]%
\begin{tabular*}{200pt}{@{\extracolsep\fill}ll@{\extracolsep\fill}}
\textbf{Parameters}&\textbf{Defintion}\\
$A_p$ & Ambiguity function\\
$E$ & Energy\\
e & Exponential function\\
$f$ & Frequency\\
$F_0$ & Frequency duration\\
$h$ & Impulse response\\
$h_n$ & Hermite function\\
$H$ & Transfer function\\
i & Complex variable\\
$i,k,l,m$ & Natural variables\\ 
$L_F$ &  Filter length  \\
$n$ & Discrete time\\
$N_B$ & Number of blocks\\
$r$ & Roll-off factor\\
$t$ & Continuous time\\
$T_c$ & Coherence time\\
$T_0$ & Signaling duration\\
$T_{CP}$ & Cyclic prefix duration\\
$T_{ZP}$ & Zero padding duration\\
$T_{GS}$ & Guard symbol duration\\
$u$ & Real variable\\
$\tau$ & Time delay\\
$\nu$ & Doppler frequency\\
$\tau_0$  & Carriers orthogonality duration\\
$\nu_0$ & Inter-carrier distance\\
$\omega$ &  Angular frequency\\ 
$\delta$ & Standard deviation\\

\end{tabular*}
\end{table}
\end{center}

\section{Prototype Filters}\label{sec2}

In parallel transmitting data systems, each channel occupies a relatively narrow frequency band. In this case, it is tempting to avoid spectral overlap of the channels to drop inter-channel interference. However, higher signaling rates can be achieved if spectral overlap is permitted \cite{saltzberg1967performance}. In multi-carrier systems, modulated signal can be represented as a linear combination of a Gabor family $ (P,T_0,F_0) $\cite{schellmann2014fbmc}  , where $ T_0 $ is the symbol duration of the modulated signal, $ F_0 $ inter carrier spacing and $P$ the prototype filter pulse. Balian-low theorem states that no modulation scheme can satisfy complex orthogonality, Nyquist rate criterion and have a good time-frequency localization (TFL) at the same time. A straightforward proof example of the theorem is Cyclic-Prefix OFDM (CP-OFDM) scheme. The Heisenberg-Gabor uncertainty is given by:

\begin{eqnarray}
\delta_t \delta_f \geq \frac{1}{4\pi}
\end{eqnarray}

Where $\delta_t $ and $\delta_f $ are respectively the time and frequency standard deviation. it means that the product of the localization uncertainties in frequency and time must exceed a fixed constant. It is significant to note that for Gaussian pulse, the equality holds due to the symmetry property of Gaussian function. Although Gaussian pulse is optimal in the sense that it presents equal spreading in both time and frequency domain, unfortunately, its pulse does not satisfy the Nyquist inter symbol interference (ISI) and inter carrier interference (ICI) cancellation property\cite{farhang2010cosine}.

\subsection{Rectangular Filter }

The rectangular function is the prototype filter used for the conventional OFDM transmission scheme. This filter distributes uniformly the energy carried by a symbol through the time domain.  At the transmitter its duration can be enlarged so it can effectively combat inter-symbol and inter-carrier interference in time invariant multi-path channels. Its analytical expression is \cite{bracewell1986fourier} : 

\begin{eqnarray}
rect(t)=\left\lbrace 
\begin{array}{ll}
1 &  \mbox{$|t|<\frac{1}{2}$}\\
0 & \mbox{$elsewhere$}
\end{array}
\right.
\end{eqnarray}

By multiplying with a suitable displaced rectangle function, it is possible to select any segment of any function. The characteristics of the rectangular filter play a significant role in determining the resulting frequency response of the OFDM system. Specifically, the convolution with the rectangular function has the effect of smoothing the signal \cite{proakis2013digital}. The Fourier transform of rectangular filter is :

\begin{eqnarray}
RECT(\omega)=\frac{1}{\sqrt{2\pi}} \frac{\sin(\frac{\omega}{2})}{\frac{\omega}{2}}
\end{eqnarray}   

With: $\omega=2 \pi f$\\ 

The implementation of a long duration rectangular filter can be considerable even after truncation. Rectangular filter accounts for a good benchmark for comparison.  

\subsection{Raised Cosine Filter}

In data communication systems, the transmitted signal must be restricted to a certain bandwidth. This can be due to either system design constraints or regulations \cite{gentile2007digital}. Thus, the choice of a filter must serve to reduce the effective bandwidth while maintaining cancellation of time interference. Raised cosine filter, takes on the shape of a sinus cardinal pulse. This filter has the property of vestigial symmetry. This means that its spectrum exhibits odd symmetry.
The transfer function of the raised cosine filter is\cite{joost2010theory}:

\begin{eqnarray}
H_{1}(\omega)=\left\lbrace 
\begin{array}{ll}
A &  \mbox{$|\omega|<\omega_1$}\\
\frac{A}{2}(1+\cos(\pi\frac{|\omega|-\omega_1}{r\omega_c}) &  \mbox{$\omega_1\leq|\omega|\leq\omega_2$}\\
0 & \mbox{$|\omega|>\omega_2$}
\end{array}
\right.
\end{eqnarray}

With: $ \omega_1=\frac{1-r}{2}\omega_c $, $ \omega_2=\frac{1+r}{2}\omega_c $, $  A=\frac{
2\pi}{\omega_c}=T_c$ , $\omega_c$ the central pulsation and $r$ the roll-off factor.\\

The energy is calculated by integrating the module of $H_1(\omega)$  over $\mathbb{R}$.

\begin{eqnarray}
E_{H_1}= \int\limits_{-\infty}^{+\infty}|H_1(\omega)|^2d\omega=\frac{\pi^2}{\omega_c}(4-r)
\end{eqnarray}

It is common to implement raised cosine response by two identical filters, one implemented at emitter side and another at receiver side. In such cases, the response becomes a square-root raised cosine. The transfer function of the root-raised cosine filter is \cite{joost2010theory}:

\begin{eqnarray}
H_{2}(\omega)=\left\lbrace 
\begin{array}{ll}
B &  \mbox{$|\omega|<\omega_1$}\\
\frac{B}{\sqrt{2}}\sqrt{1+\cos(\pi\frac{|\omega|-\omega_1}{r\omega_c}) }&  \mbox{$\omega_1\leq|\omega|\leq\omega_2$}\\
0 & \mbox{$|\omega|>\omega_2$}
\end{array}
\right.
\end{eqnarray}

With : $ B=\sqrt{T_c}$

The total energy is:

\begin{eqnarray}
E_{H_2}= \int\limits_{-\infty}^{+\infty}|H_2(\omega)|^2d\omega=2\pi
\end{eqnarray}

The impulse response of the root raised cosine filter is assumed as a real and an even function. Applying inverse Fourier transform, one can write \cite{joost2010theory}:

\begin{eqnarray}
h_2(t)=\int\limits_{-\infty}^{+\infty} H_2(\omega)\mathrm{e}^{\mathrm{i}\omega t}d\omega
\end{eqnarray}

\begin{eqnarray}
h_2(t)=C\frac{\sin\Bigg((1-r)\pi\frac{t}{T_c}\Bigg)+4r\frac{t}{T_c}\cos\Bigg((1+r)\pi\frac{t}{T_c}\Bigg)}{\pi\frac{t}{T_c}\Bigg(1-\Bigg(4r\frac{t}{T_c}\Bigg)^{2}\Bigg)}
\end{eqnarray}

With : $ C=\frac{B}{T_c}=\frac{1}{\sqrt{T_c}}$
\\

The square-root raised cosine frequency response is depicted in \textcolor{blue}{ Fig.\ref{fig1}}.
\\

The SRRC filter characteristic can be adjusted via the roll-off parameter $r$ that belongs to the interval $[0,1]$. Depending on the value of $r$, the square-root raised cosine power spectral density (PSD) can have large or narrow main lobe and great or small side-lobes. Note that as the $r$ value increases from zero to one, the passband of the filter increases while the amplitude of the time domain ripples decreases. Moreover, to get a realizable SRRC filter, the roll-off factor must be different from to zero.

\subsection{Hermite Filter}

Works on Hermite functions, aim to find a base function with dispersion product even closer to the uncertainty limit of equation (1) while satisfying the necessary orthogonality properties. Thus, the chosen base functions must be invariant when applying the Fourier transform as with Gaussian pulses. The used Hermite functions are \cite{haas1994multiple} :

\begin{eqnarray}
h_n(t)=\mathrm{e}^{-\frac{t^2}{2}}\frac{\mathrm{d^n}}{\mathrm{d}t^n}\mathrm{e}^{-t^2}
\end{eqnarray}

Which verify:

\begin{eqnarray}
\lambda .h_n(t)=\int\limits_{-\infty}^{+\infty}h_n(f)\mathrm{e}^{\mathrm{i}ft}df
\end{eqnarray}

Where:  $ \lambda=\lambda_n=j^n \sqrt{2\pi}$
\\

If we normalized by $D_n(t)=h_n(\sqrt{2\pi t})$ we get as in \cite{haas1994multiple}:

\begin{eqnarray}
D_n(t)=\int\limits_{-\infty}^{+\infty}D_n(f)\mathrm{e}^{\mathrm{i}2\pi ft}df
\end{eqnarray}

It is also demonstrated in \cite{haas1994multiple} that the pulse shape $ h(t) $ formed by linear combination of $D_n(t) $, when $n$ is multiple of $ 4 $, also satisfies this identity:

\begin{eqnarray}
h(t)=\sum\limits_{k=0}^{Q-1}H_{4k}D_{4k}(t)
\end{eqnarray}

The $ H_{4k }$ parameters are chosen so that the orthogonality condition is satisfied. Otherwise, these parameters must lead to a null ambiguity function $ A(\tau,\nu)$ at points $ \tau=nT_0 $ and $ \nu=kF_0 $. In \textcolor{blue}{ Fig.\ref{fig1}} the $ H_{4k} $ coefficients are calculated according to  \cite{haas1994multiple} for $ Q=4 $.

\subsection{Gaussian Function }

Gaussian function constitutes an isotropic prototype function. This means transforming this function by Fourier operation yields no change in the shape. For Gaussian function, the analytical expression is given by \cite{du2007classic}: 

\begin{eqnarray}
g_\alpha(t)=(2\alpha)^{(1/4)}\mathrm{e}^{-\pi \alpha t^2}, \alpha >0
\end{eqnarray}
 
By getting the Fourier transform, it yields :
 
\begin{eqnarray}
G_\alpha(f)&=(2\alpha)^{(1/4)}\int\limits_{-\infty}^{+\infty} \mathrm{e}^{-\pi \alpha t^2} \mathrm{e}^{\mathrm{-i}2\pi ft}dt \nonumber \\\nonumber\\
&\quad =(2\alpha)^{(1/4)}\sqrt{\frac{\pi}{\pi \alpha}}\mathrm{e}^{(\mathrm{-i}\pi f)^2/ (\pi \alpha)}
\end{eqnarray}

\begin{eqnarray}
= (2/\alpha)^{(1/4)}\mathrm{e}^{\mathrm{-i}\pi (f)^2/\alpha}=g_{1/\alpha}(f)
\end{eqnarray}

It is straightforward from the above equation that the Fourier transform of a Gaussian function yields  a function with the same shape except for an axis scaling factor. It seems to be an
attractive candidate for a pulse shaping prototype function, but it is in no way orthogonal:

\begin{eqnarray}
g_\alpha(t) >0
\end{eqnarray}

\subsection{IOTA Algorithm }

The basic functions forming the transmitted signal are got by translation in time and frequency of a prototype function:

\begin{eqnarray}
x_{m,n}(t)=\mathrm{i}^{m+n}\mathrm{e}^{\mathrm{i}2\pi mF_0t}x(t-nT_0)
\end{eqnarray}

Here $n$ and $m$ express the time and frequency translation respectively. Commonly $  T_0=F_0=1/\sqrt{2}$.

In the case of OFDM, this set of functions is orthonormal and forms a Hilbertian basis. Hilbertian basis provides a powerful tool to design multi-tone systems. A Hilbertian space of square-integrable functions is defined with either standard inner product or real inner product. In\cite{le1995coded}  for real inner product case, an example of Hilbertian basis of square-integrable functions is obtained with a prototype function defined by its Fourier transform:  

\begin{eqnarray}
X(f)=\left\lbrace 
\begin{array}{ll}
\frac{A}{\sqrt{F_0}}\cos(\frac{\pi f}{2F_0}) &  \mbox{$|f|<F_0$}\\
0 & \mbox{$elsewhere$}
\end{array}
\right.
\end{eqnarray}

Consider an orthogonalization factor, denoted $ O $, which aims to orthogonalise a given function over frequency axis. The same function can be orthogonalized oven the time axis using the operator defined by $ F^{-1}OF $ ($F$ being the Fourier transform).   When applied to the function, the $ O $ operator cancels the ambiguity function over the frequency axis. Furthermore, we can see the same effect on the ambiguity function over time axis by applying the $ F^{-1}OF $ operator. Let's consider the following functions:

\begin{eqnarray}
g_{\alpha}(t)=\sqrt[4]{2 \alpha}. \mathrm{e}^{-\pi \alpha t^2}
\end{eqnarray}

\begin{eqnarray}
G_{\alpha}=F^{-1}OFOg_{\alpha}(t)
\end{eqnarray}

It can be demonstrated that:

\begin{eqnarray}
\begin{array}{ll}
A_{G_{\alpha}}(n\sqrt{2},m\sqrt{2})=0 & \mbox{$(m,n)\neq(0,0)$}
\end{array}
\end{eqnarray}
 Where $A_{G_{\alpha}}$ is the ambiguity function described in equation (38).\\
  
Here we have a set of functions that can be taken as the prototype function of a Hilbertian basis depending on the chosen term $ \alpha $. Moreover, this parameter affects the power distribution of the pulse in the time/frequency space\cite{schellmann2014fbmc}. It is important to note that the $ O $ operator is defined to satisfy (23). The first orthogonalization step introduces nulls in the ambiguity function in frequency domain while the second orthogonalization step introduces nulls in delay domain \cite{farhang2011ofdm}.

\begin{eqnarray}
y(u)=\frac{2^{\frac{1}{4}}x(u)}{\sqrt{\sum\limits_{k}\| x(u-\frac{k}{\sqrt{2}})\|^2}}
\end{eqnarray}

\subsection{Extended Gaussian Function }

It is shown in \cite{roche1997family} that an analytical expression can be derived for the isotropic orthogonal transform algorithm (IOTA) algorithm given by equation (21). This makes the function $G_\alpha(t)$ in this equation more tractable:

 \begin{eqnarray}
G_{\alpha,\tau_0,\nu_0}(t)&=\frac{1}{2}\sum\limits_{k=0}^{\infty}d_{k,\alpha,\nu_0}\left[g_{\alpha}\left( t+(\frac{k}{\nu_0})\right)+g_{\alpha}\left( t-(\frac{k}{\nu_0})\right) \right] \nonumber\\
&\quad \times \sum\limits_{l=0}^{\infty}d_{l,\tau_0,1/\alpha}\cos\left(2\pi l \frac{t}{\tau_0}\right)
\end{eqnarray}

With $d_{k,\alpha,\nu_0}$ a set of real coefficients and $ 0.264< \alpha<1/0.264$. 

The practical use of the above equation needs the knowledge of the real coefficients $d_{k,\alpha,\nu_0}$. These are expressed by the following infinite series:

\begin{eqnarray}
d_{k,\alpha,\nu_0}=\sum\limits_{l=0}^{\infty}a_{k,l}\mathrm{e}^{-(l\pi/2\nu_0^2 )\alpha}
\end{eqnarray} 

We note that we have to compute an infinite set of coefficients $a_{k,l}$. However, it is still possible to truncate the above expression to a finite set with  enough  accuracy, due to the fast decreasing of the function $G_{\alpha}$ \cite{siohan2000cosine} . A set of $a_{k,l}$ coefficient is reported in \cite{roche1997family}witch can be used in calculating the  seven first $d_{k,\alpha}$ parameters.

\subsection{Martin Filter }
Martin proposes in \cite{martin1998small}  a linear phase filter based on Lerner filters. The idea is such that a linear combination is performed on a number of Inverse Fourier Transformation/ Fourier Transformation (IFT/FT) adjacent filter channels where the sum of weighting coefficients is fixed to be zero with an alternation in coefficient's sign:
\begin{eqnarray}
T_i(\omega)=\sum\limits_{i}k_i \mathrm{IFT}_i=...+k_{i-2} \mathrm{IFT}_{i-2}  ++k_{i-1} \mathrm{IFT}_{i-1} \nonumber\\ 
+k_{i} \mathrm{IFT}_i+k_{i+1} \mathrm{IFT}_{i+1}+k_{i+2} \mathrm{IFT}_{i+2}+...       
\end{eqnarray}

With:
\begin{eqnarray}
\sum\limits_{i}k_i=0      
\end{eqnarray} 

 This constraint results in an excellent stop-band performance. The choice of coefficients is paramount for determining the adaptation to a given application scheme. If some choices result for example, in an aliasing for multi-tone communications, others can be acceptable. In \cite{martin1998small} filter coefficients are determined for the case where $M$ the number of filter banks, satisfies:
   
\begin{eqnarray}
M=\left\lbrace 
\begin{array}{ll}
n/3 &  \\
n/4 &  \\
n/6 & \\
n/8 & \\
\end{array}
\right.
\end{eqnarray}

Here, $n$ is the order of IFT.
In the case $M=n/3$, five adjacent filters are used, the order of the resulting prototype filter is $n-1$ and the number of weighting coefficients is  $(2n/M)-1$ \cite{martin1998small}: 

\begin{eqnarray}
T_{i/3}(\omega)=k_{2} \mathrm{IFT}_{i-2}+k_{1} \mathrm{IFT}_{i-1}+k_{0} \mathrm{IFT}_i\nonumber\\ 
+k_{1} \mathrm{IFT}_{i+1}+k_{2} \mathrm{IFT}_{i+2}       
\end{eqnarray}
Where:

\begin{eqnarray}
\mathrm{IFT}_{i}(\omega)=\frac{1}{n}\sum\limits_{k=0}^{n} \mathrm{e}^{-\mathrm{j}2\pi i k/n} \mathrm{e}^{-\mathrm{j}\omega k T}      
\end{eqnarray}

Under the constraint that: 

\begin{eqnarray}
k_{2} +k_{1} +k_{0} +k_{1} +k_{2}=0       
\end{eqnarray}

Generally, other considerations are taken into account to find filter coefficients, as the unity gain for the sum of transmitting and receiving filters at the IFT center frequencies:  

\begin{eqnarray}
\sum\limits_{i=0}^{K} T_i(\omega)R_i(\omega)=1
\end{eqnarray}

\onecolumn
\begin{figure}
\centerline{\includegraphics[width=500pt,height=50pc]{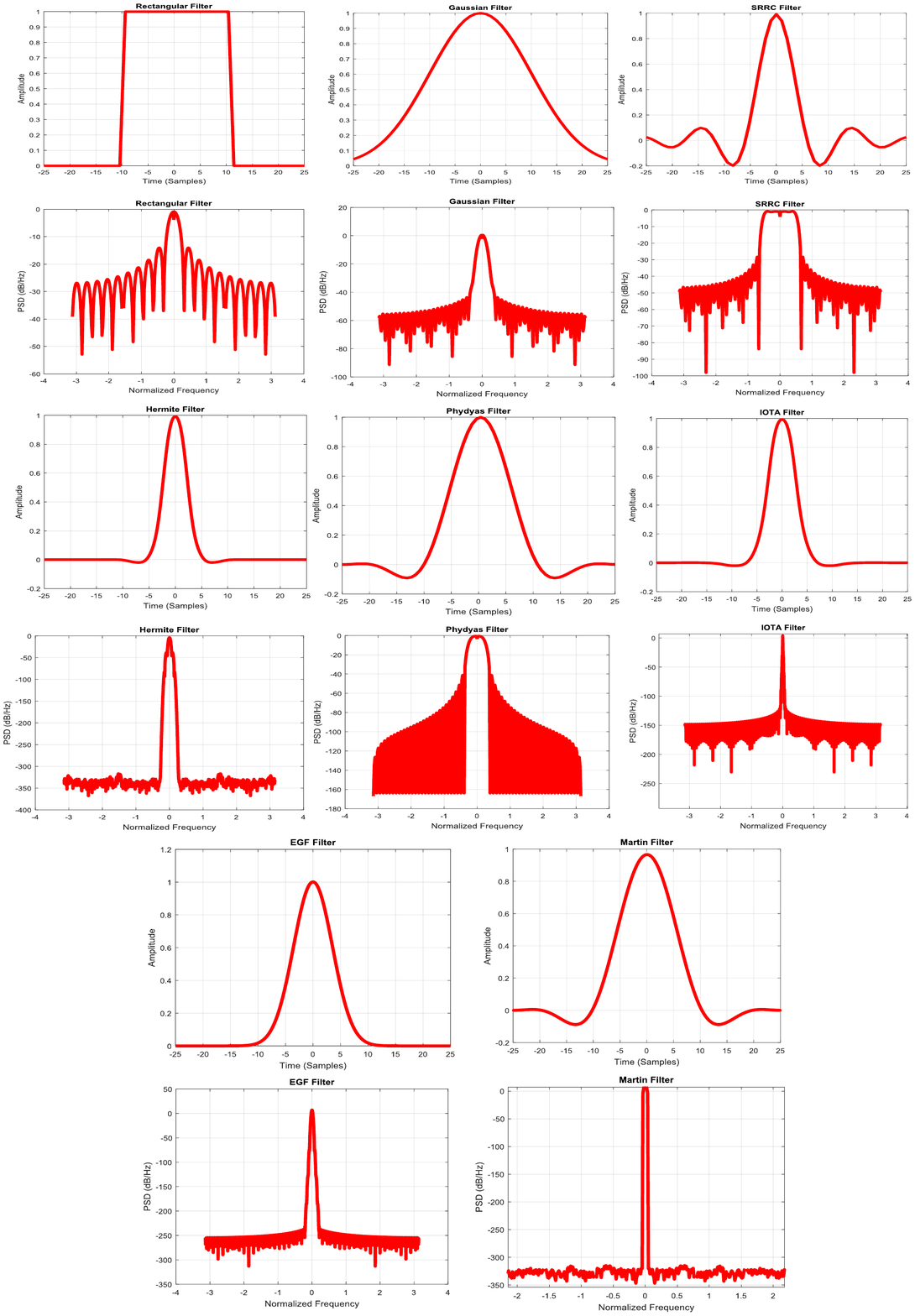}}
\caption{Time-Frequency representation of Prototype filters\label{fig1}}
\end{figure}

\begin{center}
\begin{table}
\centering
\caption{Analytical expression of prototype filters.\label{tab1}}
\begin{tabular*}{560pt}{|l|cccc}
\hline
\textbf{Filter} & \textbf{Alalytical Expression} & \textbf{Parameters}  \\\\
\hline\\
Rectangular & $ h(t)=\left\lbrace 
\begin{array}{ll}
1 &  \mbox{$|t|<\frac{1}{2}$}\\
0 & \mbox{$elsewhere$}
\end{array}
\right.$\\\\
SRRC & $
h(t)=C\frac{\sin\Bigg((1-r)\pi\frac{t}{T_c}\Bigg)+4r\frac{t}{T_c}\cos\Bigg((1+r)\pi\frac{t}{T_c}\Bigg)}{\pi\frac{t}{T_c}\Bigg(1-\Bigg(4r\frac{t}{T_c}\Bigg)^{2}\Bigg)}$ & $C=\frac{B}{T_c}=\frac{1}{\sqrt{T_c}}, r[0,1]$   \\\\
Hermite & $h_n(t)=\mathrm{e}^{-\frac{t^2}{2}}\frac{\mathrm{d^n}}{\mathrm{d}t^n}\mathrm{e}^{-t^2}$ & $Q>0$  \\\\

Gaussian & $g_\alpha(t)=(2\alpha)^{(1/4)}\mathrm{e}^{-\pi \alpha t^2}$&  $\alpha >0$ \\\\
 

IOTA & $h_{\alpha}(t)=F^{-1}OFOg_{\alpha}(t)$ &$y(u)=\frac{2^{\frac{1}{4}}x(u)}{\sqrt{\sum\limits_{k}\| x(u-\frac{k}{\sqrt{2}})\|^2}}$  \\\\
EGF& $G_{\alpha,\tau_0,\nu_0}(t)=\frac{1}{2}\sum\limits_{k=0}^{\infty}d_{k,\tau_0,\nu_0}\left[g_{\alpha}\left( t+(\frac{k}{\nu_0})\right)+g_{\alpha}\left( t-(\frac{k}{\nu_0})\right) \right].\sum\limits_{l=0}^{\infty}d_{l,\tau_0,1/\alpha}\cos\left(2\pi l \frac{t}{\tau_0}\right)$&$d_{k,\alpha,\nu_0}=\sum\limits_{l=0}^{\infty}a_{k,l}\mathrm{e}^{-(l\pi/2\nu_0^2 )\alpha}$        \\\\


MARTIN & $
T_i(\omega)=..+k_{i-1} \mathrm{IFT}_{i-1}+k_{i} \mathrm{IFT}_i+k_{i+1} \mathrm{IFT}_{i+1}+.. $& $\mathrm{IFT}_{i}(\omega)=\frac{1}{n}\sum\limits_{k=0}^{n} \mathrm{e}^{-\mathrm{j}2\pi i k/n} \mathrm{e}^{-\mathrm{j}\omega k T}$   \\\\

PHYDYAS & $h(t)=1+2\sum\limits_{k=1}^{K-1}H_k \cos \left(2 \pi \frac{kt}{Kt} \right)$  & $K=2,3,4$ \\\\
\hline
\end{tabular*}
\end{table}
\end{center}

\begin{center}
\begin{table}
\centering
\caption{Papers dealing with Prototype Filters.\label{tab2}}
\begin{tabular*}{550pt}{|l|ccccccccccccccc}
\hline\\
\textbf{Filters} & \textbf{Rectangular} && \textbf{SRRC}&&\textbf{Hermite}&&\textbf{Gaussian}&&\textbf{IOTA}&&\textbf{EGF}\ && \textbf{MARTIN}&&\textbf{PHYDYAS}\\\\
\hline\\\\
\textbf{Equalization} & \cite{stitz2010pilot}, \cite{baltar2010mlse} && \cite{cherubini2002filtered}, \cite{ihalainen2011channel} && \cite{aldirmaz2010spectrally}&& \cite{rusek2009multistream}, \cite{sahin2013investigation}&&\cite{lele2008channel}, \cite{lele2007preamble} &&  && \cite{ikhlef2009enhanced}, \cite{baltar2010mlse}&& \cite{ikhlef2009enhanced}, \cite{zakaria2010maximum}\\
& \cite{sahin2013investigation}, \cite{falconer2002frequency} && \cite{stitz2008practical} && && && &&  && \cite{stitz2010pilot}, \cite{bellanger2008filter}&&\\\\
\textbf{Synchronization}& \cite{fusco2008sensitivity},\cite{ringset2010performance} && \cite{saeedi2011sensitivity}, \cite{fusco2008sensitivity} && \cite{saeedi2011sensitivity} && && \cite{saeedi2011sensitivity}&& \cite{du2008pulse1}, \cite{du2008pulse} && \cite{ikhlef2009enhanced}, \cite{schaich2010filterbank} && \cite{schaich2010filterbank}\\
& \cite{du2008pulse} && \cite{fusco2007blind} && && && && && \cite{ringset2010performance}, \cite{fusco2009joint} &&\\\\
\textbf{Complexity Analysis} & \cite{zhang2008oversampled}, \cite{baltar2007out} && \cite{waldhauser2008mmse}, \cite{waldhauser2008adaptive} && && && \cite{mehmood2012hardware}&& && \cite{bellanger2008filter}&& \cite{bellanger2008filter} \\
& \cite{waldhauser2009adaptive} &&  && && && &&  &&  &&[1],[2]\\\\
\textbf{Channel Estimation} & \cite{du2009novel},\cite{lele20082} && \cite{du2008pulse1}, \cite{han2007hexagonal}&& && && \cite{javaudin2003pilot}, \cite{lele20082}&& \cite{du2008pulse1},\cite{lele20082} &&  && \cite{kofidis2013preamble}\\
&  && \cite{du2009novel} && && && \cite{lele2007preamble}&& \cite{du2009novel} &&  &&\\\\
\textbf{Spectral Leakage} & \cite{kollar2011physical}, \cite{baltar2007out} && \cite{baltar2007out}, \cite{zhang2008oversampled} && && && \cite{alard1996construction}&& && \cite{jiang2012suppressing}&& \\
& \cite{zhang2008oversampled} && \cite{ihalainen2006channel}&& && && &&  && && \\\\
\textbf{PAPR} & \cite{ihalainen2009filter},\cite{waldhauser2006comparison} && \cite{waldhauser2006comparison}, \cite{ihalainen2009filter} && && && &&  && \cite{viholainen2009deliverable} &&\cite{viholainen2009deliverable}\\
& \cite{ihalainen2006channel} &&  && && && &&  &&  && \\\\
\hline
\end{tabular*}
\end{table}
\end{center}
\twocolumn

Where: $T_i(\omega)$ are the transmission filters and $R_i(\omega)$ the received ones.
 
Witch yields:

\begin{eqnarray}
\left\lbrace 
\begin{array}{ll}
k_{0}=1\\
k_{1}^2+k_{2}^2=1
\end{array}
\right.
\end{eqnarray}

\subsection{Phydyas Filter  }
The design of the PHYDYAS prototype filter is based on a frequency sampling technique. The Nyquist theory imposes that the  impulse response of the transmission
filter must cross the zero axis at all the integer multiples of the symbol period. In frequency domain, this
translates to a symmetry condition about the cutoff frequency. So the design is based on considering the frequency coefficients while imposing symmetry condition \cite{bellanger2010fbmc}. The frequency response of the prototype filter is obtained from the frequency coefficients using an interpolation function:

\begin{eqnarray}
H(f)=\sum\limits_{k=-K+1}^{K-1} H_k \frac{\sin\left(\pi(f-\frac{k}{MK})MK\right) }{MK\sin\left(\pi(f-\frac{k}{MK})\right)} 
\end{eqnarray}

$K$ represents the overlapping factor and $M$ is the number of sub-channels. In time domain, this yields:

\begin{eqnarray}
h(t)=1+2\sum\limits_{k=1}^{K-1}H_k \cos \left(2 \pi \frac{kt}{Kt} \right)
\end{eqnarray}
   
For a discrete time, a close form of the prototype filter for an overlapping factor $K=4$ is expressed as: 

\begin{eqnarray}
h[m]=H[0]+2\sum\limits_{k=1}^{K-1} (-1)^k H[k] \cos \left( \frac{2\pi k}{KM}(m+1) \right)
\end{eqnarray}
Where:

\begin{eqnarray}
\left\lbrace 
\begin{array}{ll}
 m=0,1,...,KM-2\\
 K=4\\
 H[0]=1\\
 H[1]=0.97195983\\
 H[2]=1/\sqrt{2}\\
 H[3]=\sqrt{1-H[1]}
\end{array}
\right.
\end{eqnarray}

The analytical expressions of prototype filters are shown in \textcolor{blue}{Table.\ref{tab1}}.

\section{ Selection Criteria }\label{sec2}

\subsection{Ambiguity Function Analysis}

The orthogonality condition can be investigated through ambiguity function. For a pulse shape $ p(t) $, we define his ambiguity function as\cite{stein1981algorithms} : 

\begin{eqnarray}
A_p(\tau,\nu)=\int
\limits_{-\infty}^{+\infty}p(t+\frac{T}{2})p^*(t-\frac{T}{2})\mathrm{e}^{-\mathrm{i}2\pi \nu t}dt
\end{eqnarray}

Where $ \tau $ is the time delay and  $\nu$ is the frequency Doppler. In this section, we will examine the set of ambiguity functions calculated for the already studied filters.
\\ 

The ambiguity function indicates how symbols are spread across the time and frequency axes; a zero crossing of the ambiguity function over time or frequency axis guarantees ISI and ICI free transmission. \textcolor{blue}{ Fig.\ref{fig2}} presents the ambiguity surface for different filter designs. For square root raised cosine filter with a roll-off factor $ \alpha =0.1 $ the ambiguity function value attains $-10$dB ($0.1$ on color bar) at the points $\nu=\pm20$ KHz and $\tau=\pm 70\mu$s. The ambiguity function reaches a level lower than $-50$dB (dark blue on the \textcolor{blue}{ Fig.\ref{fig2}}) as $\nu \geq \pm 35$KHz and $\tau \geq \pm 180\mu$s. From the ambiguity surface for the Gaussian filter, we note that the symbol spreading effect over time and frequency equals. This is due the symmetry of the Gaussian function with respect to the Fourier transform. From the \textcolor{blue}{ Fig.\ref{fig2}} , one can conclude that in terms of orthogonality, the Gaussian function is not a good choice since it is non-orthogonal and shows a great spreading effect across the time-frequency axes. The IOTA design algorithm presented is based on two orthogonalization parameters; the first introduces nulls in the ambiguity function in frequency domain, while the second introduces nulls in the ambiguity function in the time domain. The space in the figure characterizes by $\nu \geq \pm10$KHz and $\tau\geq \pm 250\mu$s are areas where $|A_p |\leq-60$dB.  The Hermite pulse ambiguity function is plotted  with parameters $H_{4k }=\lbrace H_0, H_4, H_8, H_{12} \rbrace $ where $H_0=1.1850899$. The Hermite pulses suffer from numerical problems, thus its ambiguity function calculation takes a long time in the simulation. It is clear that ambiguity function reaches swiftly lower values as $\nu=\pm15$KHz; nevertheless, we do not see the same case on a time axis. As for the ambiguity surface of a Phydyas filter with parameters $ H_1=0.97196$,  $H_2=0.70710$  and $H_3=0.23514$, like previous ambiguity function scheme, it presents high values around $\nu\approx0$KHz and $\tau \approx0\mu$s point, meanwhile it shows some regions near the origin where it decreases rudely as the SRRC, IOTA and Hermite cases. The Martin filter ambiguity function is plotted for the case $K=6$, time and doppler delay are relatively high for region inside $\tau<\pm 50\mu$s  and  $\nu<\pm15$KHz. The ambiguity function reaches low values outside this region. Finally, Regarding EGF filter, ambiguity function exhibits same behavior in time and frequency domain since it is just an extension of Gaussian filter, nevertheless it reaches promptly very low values $|A_p |\leq-60$dB as can be deduced.


\begin{figure}
\centerline{\includegraphics[width=250pt,height=20pc]{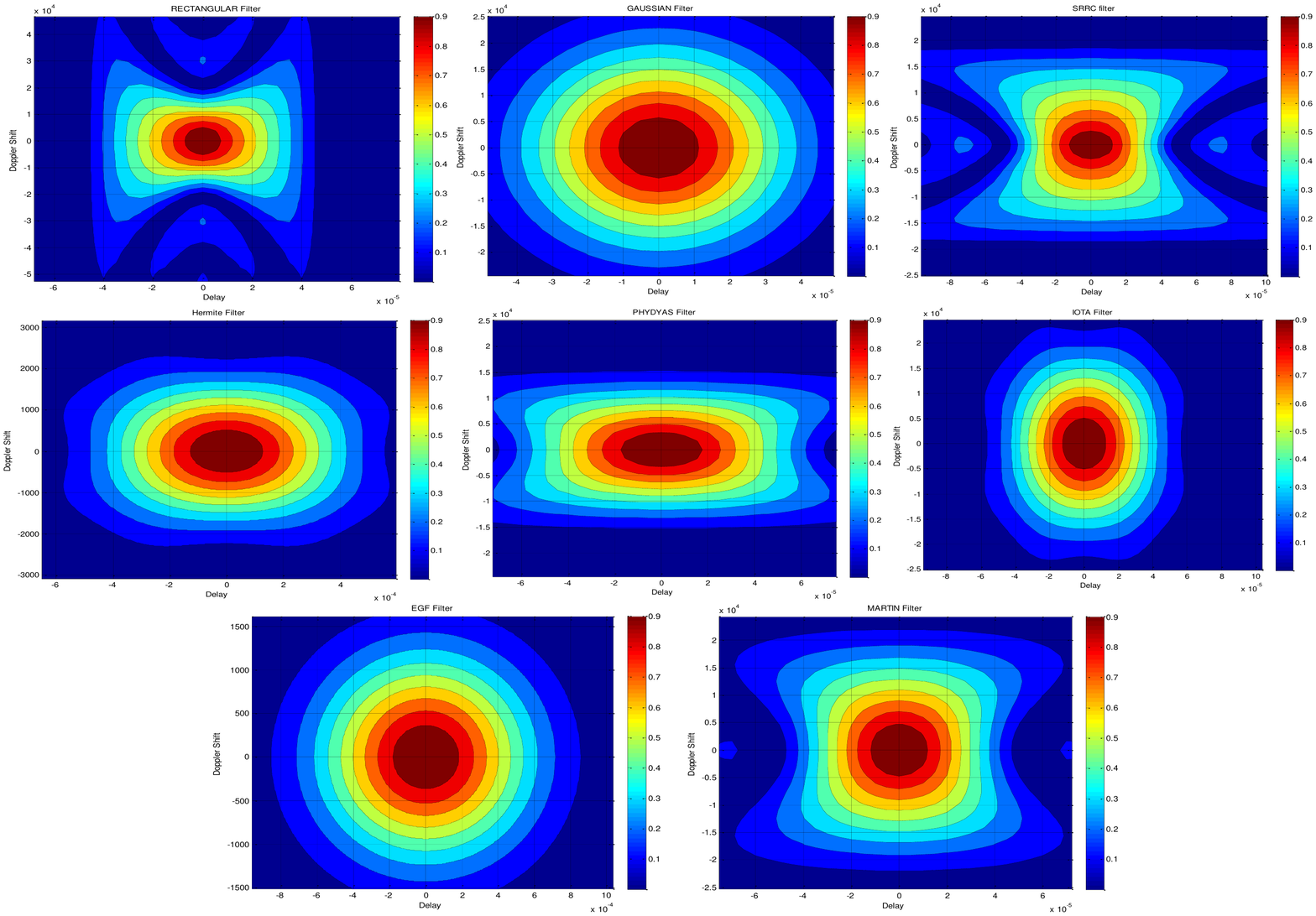}}
\caption{Ambiguity function.\label{fig2}}
\end{figure}

\subsection{Energy Concentration }

In practice, it is always preferable to limit the duration of the pulse shape so that the computational complexity is significantly reduced. However, limiting pulse shape by truncation, for instance, causes high side-lobes in Fourier domain. So How can we obtain a time-limited pulse with least out-of-band leakage or inversely a band-limited pulse with  greatest energy concentration within a given interval? The selected prototype filter depending on energy concentration should discuss this trade-off problem. \textcolor{blue}{ Fig.\ref{fig3}}, presents a comparison of all mentioned filters in terms of energy concentration. The rectangular filter energy is shown to be spread out on all frequency domain. This finding is nothing but a direct result of the Balian-Low theorem, which states that no scheme can satisfy simultaneously a good time-frequency localization (TFL). The Gaussian function is another filter with poor energy concentration. This function was subject in several studies to enhance his energy localization and reducing out-of-band leakage. The resulting filters are IOTA and EGF. The bandwidths of these filters at $-100$dB power level are $5$Hz and $12$Hz, respectively . The EGF is plotted with $\alpha$ parameter equals $1/2$. It is important to note that an increase in the value of $\alpha$ improves spectral efficiency of the filter. The SRRC filter shows a good energy concentration at the expense of a nearly constant side-lobes, its bandwidth at $-100$ is $10$Hz. Martin and Hermite filters exhibit large main lobe with a bandwidth of $10$Hz and $11$Hz at power levels $-70$dB and $-80$dB, respectively. As for Phydias filter, its main lobe bandwidth is about $60$Hz at $-70$dB.      

\begin{figure}[!b]
\centerline{\includegraphics[width=300pt,height=20pc]{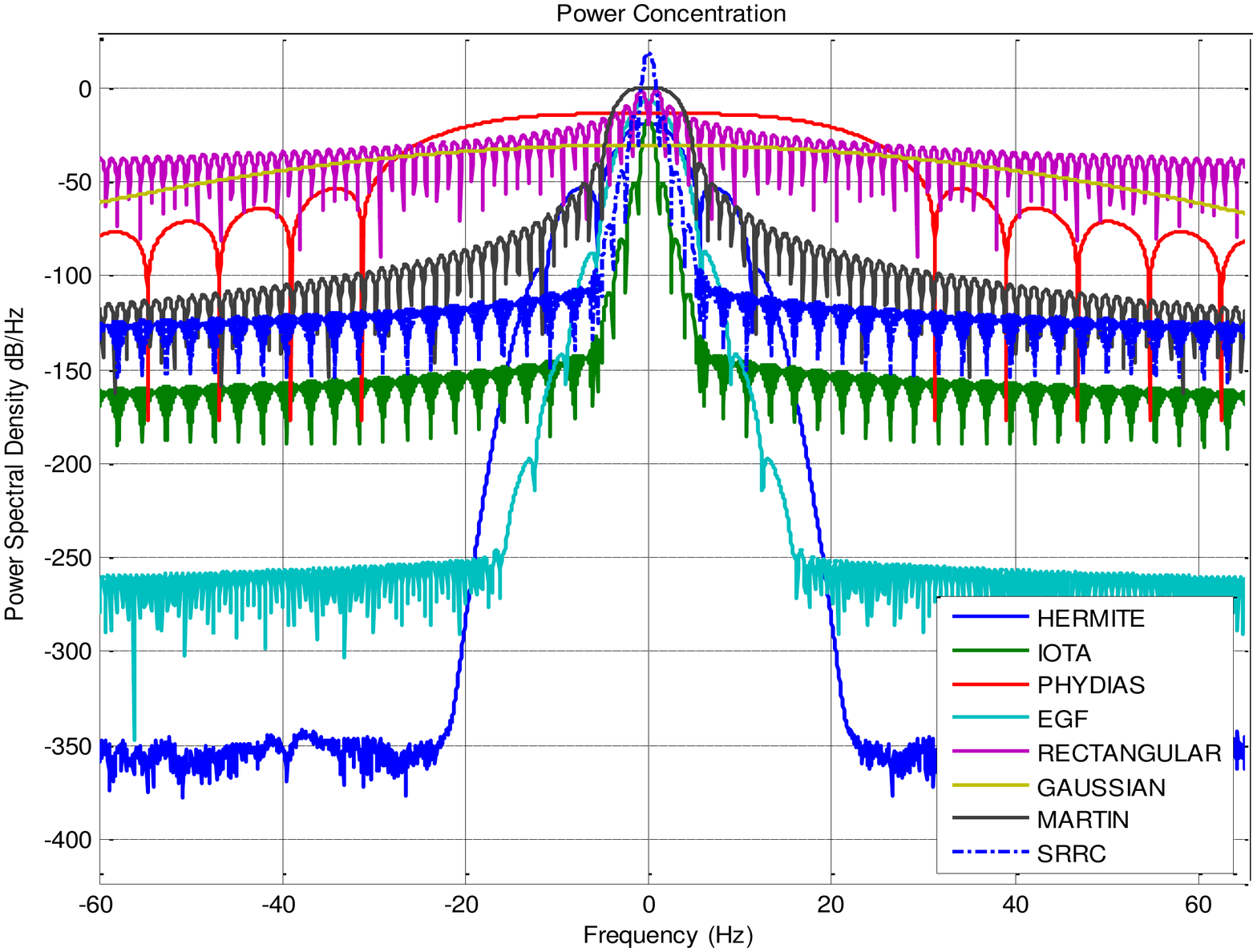}}
\caption{Energy concentration.\label{fig3}}
\end{figure}

\subsection{Transition Band Decay }

In many applications, it is preferable for a given transmitting model that the  side-lobes decay rapidly rather than remaining constant. The motivation behind this is to reduce considerably interference among adjacent users. It is known that the side-lobes falling rate is related to the number of continuous derivatives, this one is nothing but a filter's smoothing measure \cite{harris1978use} \cite{singla2009desired}. \textcolor{blue}{ Fig.\ref{fig4}}, shows the decaying property for the different presented prototype filters in the frequency domain. Due to the well localization of the rectangular filter in the time domain, it has no significant decay in the frequency domain and the side-lobes remain constant around $-40$dB. Gaussian filter shows also a slow decay until $120$Hz, then the side-lobes remain nearly constant around $-160$dB. The Phydias and SRRC filters evince a similar decay behavior with a difference of about $20$dB that increases with frequency, since SRRC side-lobes come to be constant with frequency. For the Martin filter, the coefficient calculation leads to continuous derivatives of the function which ensure a magnificent falling property. The continuous decay of side-lobes is well established in the \textcolor{blue}{ Fig.\ref{fig4}}. IOTA, EGF and Hermite filters show the fast decaying property among all filters. The PSD of the three filters reaches a very low thresholds about $-160$dB only for the first $20$Hz band. Depending on the chosen $\alpha$ coefficient, the EGF filter exhibits roughly good results. IOTA is a particular EGF filter derived from a mathematical closed-form expression with $\alpha=1$. For Hermite filter the slope characteristic of the PSD can be further enhanced by increasing the filter order $Q$ at the expense of an increase in complexity.

\section{Filtered waveforms}\label{sec2}

Considering the needs of upcoming applications in wireless communications, new filtered waveforms have been studied to be adopted in physical communication layer. These ones include several novel prototype filters which provide a good frequency localization and reduce the out-of-band radiation. Some techniques as Filter-Bank Multi-Carrier Offset QAM (FBMC-OQAM) and GFDM try to filter each sub-carrier separately to get a tight time-frequency localization in each sub-carrier, while others like F-OFDM apply the filter to the whole occupied band  or to groups of consecutive sub-carriers as UFMC technique to reduce the Out-of-band (OOB) leakage. 

\begin{figure}
\centerline{\includegraphics[width=300pt,height=20pc]{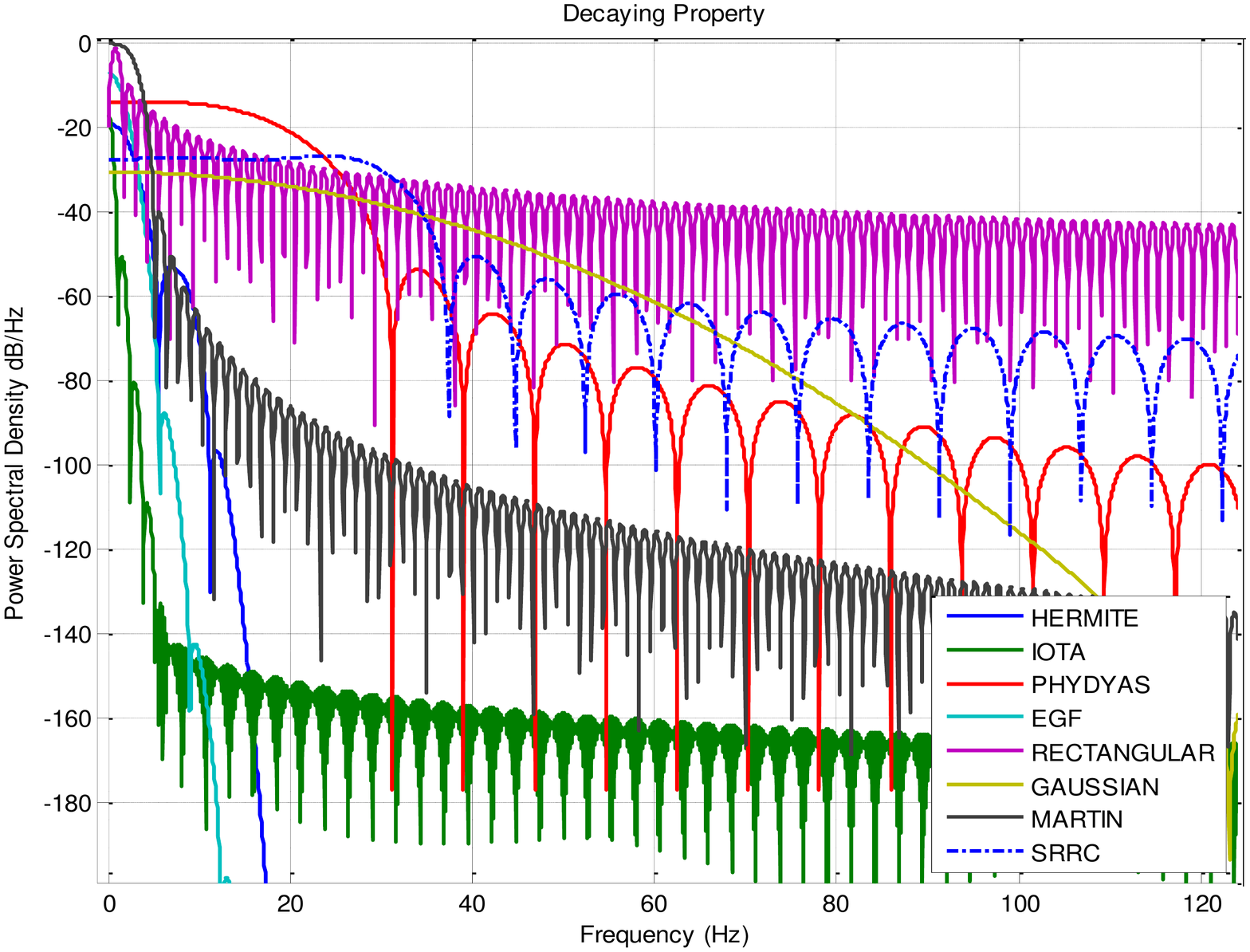}}
\caption{Decaying property.\label{fig4}}
\end{figure}

\begin{figure}[!b]
\centerline{\includegraphics[width=280pt,height=20pc]{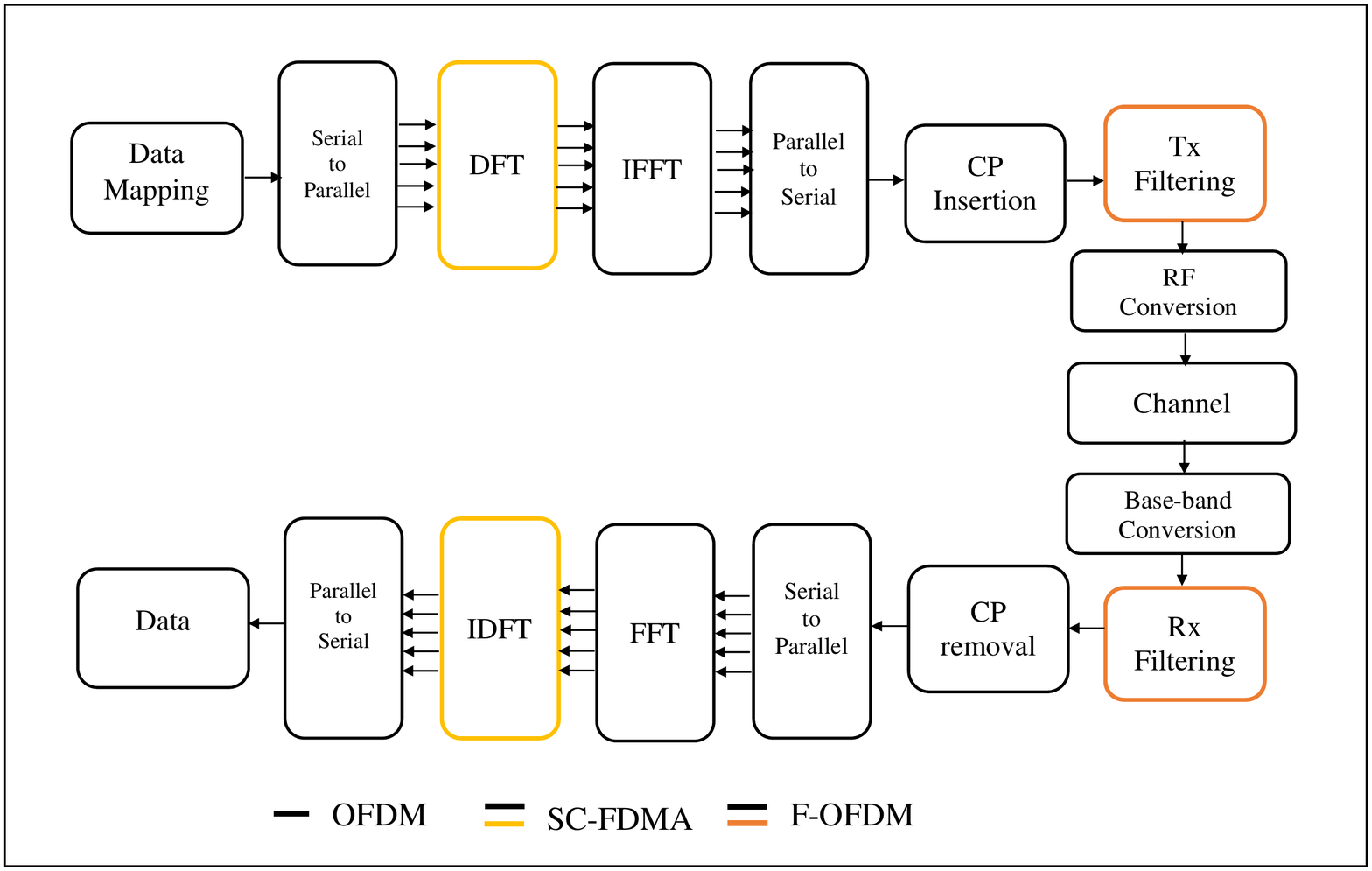}}
\caption{CP-OFDM, SC-FDMA and F-OFDM tranceivers.\label{fig5}}
\end{figure}

\subsection{CP-OFDM}

The single-carrier frequency division multiple access (SC-FDMA) and CP-OFDM techniques are used respectively in the up-link and down-link of the actual LTE systems. SC-FDMA has the characteristic to improve the energy efficiency at the User Equipment (UE) through its low PAPR, this feature is due to  an added stage of pre-coding (DFT) that is introduced in the OFDM chain (before IFFT). The linking  between the DFT and IFFT stages can be regarded as a single carrier signaling that gets at lower PAPR by comparison to multi-carrier modulations. The CP-OFDM technique uses several sub-carriers to send a set of parallel low rate symbols. These complex symbols take values from a given constellation scheme. A cyclic prefix is added to each OFDM symbol as a guard interval to cope with the channel distortion. OFDM technique shows a low complexity in implementation, at the same time it achieves a good bandwidth efficiency. The simplicity of implementation is a consequence of the fact that each OFDM symbol is a sum of pure tones. Moreover, the equalization is ensured by only one tap equalizer. OFDM is no more available when the transmission is carried out on non-contiguous bands or when the channel is double dispersive. This because OFDM uses a square window in time domain and requires stringent time synchronization, its adoption in such environments has been found very difficult to set up. The time expression of the OFDM is given by equation (39) and the operating principle is shown in \textcolor{blue}{ Fig.\ref{fig5}}.

\begin{eqnarray}
x_{OFDM}(t)= \frac{1}{\sqrt{N}}\sum\limits_{k=0}^{N-1} \Bigg \lbrace x_k \times \Pi(t)\times e^{-\mathrm{i}2\pi f_kt} \Bigg\rbrace
\end{eqnarray}

 Where $x_k$  denoted the complex symbols that emanate from a given constellation (e.g: QAM),  $\Pi(t)$ denote the rectangular waveform filter and $N$ the number of sub-carriers. This equation can be further developed to be : 
 \begin{eqnarray}
=(-1)^n \times TFD^{-1} \Bigg\lbrace x_k \sqrt{N} \Bigg\rbrace_{k=0}^{N-1} 
\end{eqnarray}

\begin{figure}
\centerline{\includegraphics[width=250pt,height=20pc]{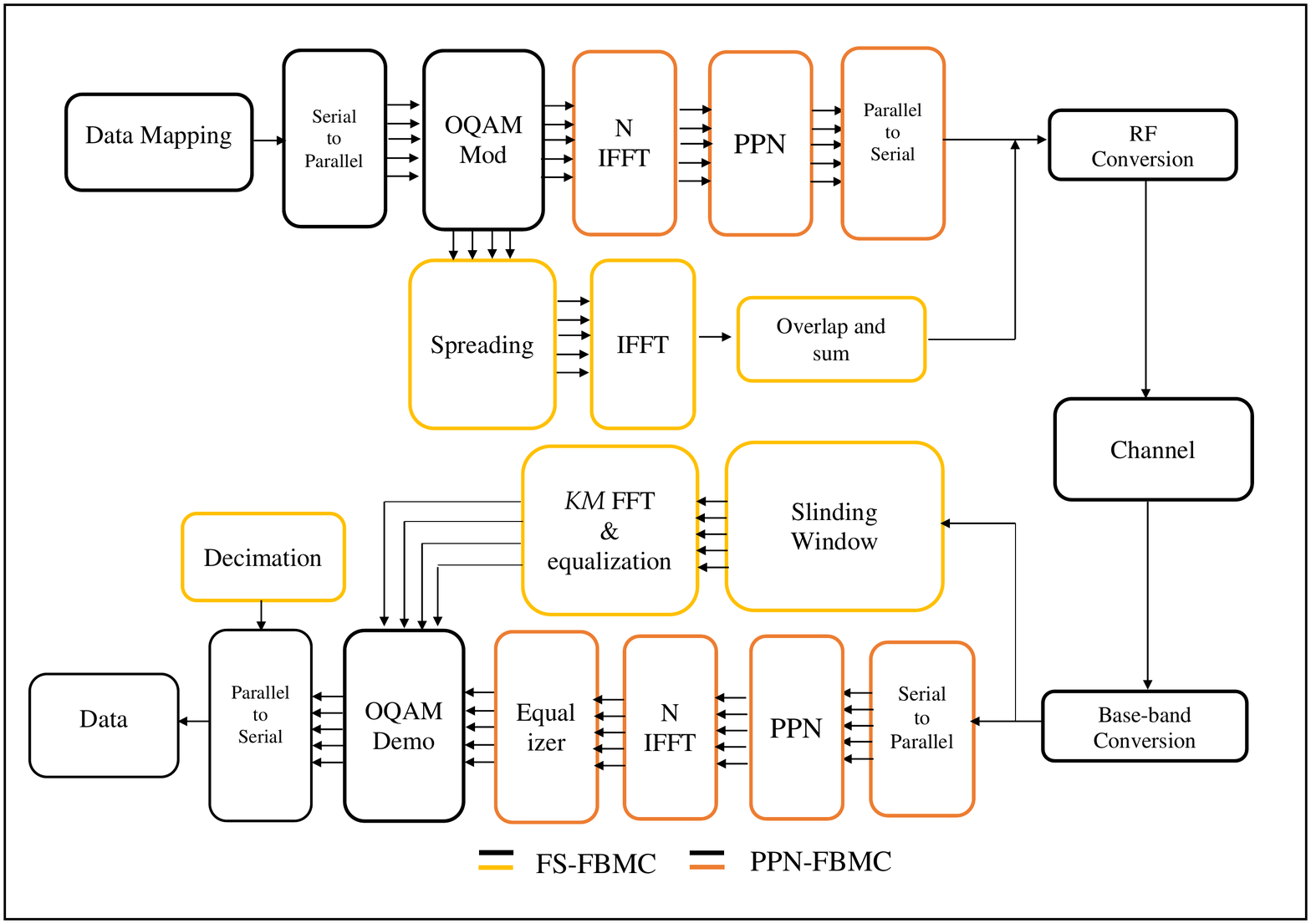}}
\caption{PPN-FBMC and FS-FBMC tranceivers.\label{fig6}}
\end{figure}
  
\subsection{FBMC}

The filter bank multi-carrier (FBMC) modulation was first introduced by Chang \cite{chang1966synthesis} and Saltzberg \cite{saltzberg1967performance}. The technique consists of  filtering each sub-carrier at both emitter ( synthesis filter) and receiver ( analysis filter) in order to address the spectral OOB leakage of OFDM. The length of the prototype filter is generally $K$ times the number of sub-carriers ( size of the FFT), where $K$ is the overlapping factor. Thus, each FBMC symbol overlaps with $K$ neighboring symbols in the time domain. The prototype filter must satisfy the Nyquist criterion to steer clear of the inter-symbol interference (ISI).The FBMC variants are \cite{farhang2010cosine}: Cosine modulated multi-tone scheme (CMT) where a set of pulse amplitude modulation (PAM) symbols are transmitted using vestigial side band (VSB) sub-carrier channels, the filtered multi-tone (FMT) based on complex QAM signaling and the staggered multi-tone scheme (SMT) based on double sideband (DSB) modulation, also referred to (FBMC/OQAM) for real valued offset QAM. Counter to FMT technique, FBMC-OQAM is able to achieve a maximum spectral efficiency (SE) due to its real orthogonality. So, OQAM does not exhibit any flaw  compared to QAM since they achieve the same SE. Nevertheless, some techniques developed for CP-OFDM like pilot-based channel estimation and adaptation to Multiple-Input Multiple-Output (MIMO) systems are no more applicable to FBMC-OQAM. The FBMC structure can be efficiently implemented using IFFTs or FFTs combined with a polyphase network (PPN) \cite{bellanger2010fbmc} or using a frequency spreading (FS) approach \cite{bellanger2012fs} proposed recently as an alternative to PPN approach. The PPN-FBMC modulator shown in \textcolor{blue}{ Fig.\ref{fig6}}, consists of an OQAM modulator, an IFFT of order $N$ and polyphase networks bloc. The receiver applies a matched filter then performs an FFT of size $N$, the result feeds a multi-tap equalizer. In FS-FBMC (see \textcolor{blue}{ Fig.\ref{fig6}}), OQAM symbols are filtered in frequency domain before the extended IFFT bloc by order $KN$. At the receiver side, a sliding window selects $KN$
points every $N/2$ samples. Then, a FFT of size $KM$
is applied followed by a filtering operation. The transmitted FBMC-OQAM signal is expressed as follows: 

\begin{eqnarray}
x_{FBMC}(t)= \sum\limits_{k=0}^{K-1} \sum\limits_{m=0}^{M-1}\Bigg \lbrace d_{k,m}\theta_{k,m} \times g(n-mK/2)\times e^{ \frac{\mathrm{i}2\pi kn}{K}}  \Bigg\rbrace
\end{eqnarray}

Where $K$ and $M$ are respectively the number of sub-carriers and the number of symbols, $d_{k,m}$ is the corresponding data symbol to be transmitted within the $k^{th}$ frequency sub-carrier and the $m^{th}$ time symbol, $g(n)$ refers to the prototype filter time coefficients (e.g.: IOTA\cite{mehmood2012hardware}, PHYDIAS\cite{jeon2016prototype}, etc.). The $\theta_{k,m}$ term is defined as:   

\begin{eqnarray}
\theta_{k,m}=\left\lbrace
\begin{array}{ll}
\pm 1, &  \mbox{ if $m+k$ is even,}\\
\\
 \pm \mathrm{j}, &  \mbox{ if $m+k$ is odd,}
\end{array}
\right.
\end{eqnarray}
 
A design and implementation of FBMC-OQAM transceiver on Field Programmable Gate Arrays (FPGA) logic device is presented in \cite{shaheen2019design} and \cite{keerthana2019fpga}. 

\begin{figure}[!b]
\centerline{\includegraphics[width=250pt,height=20pc]{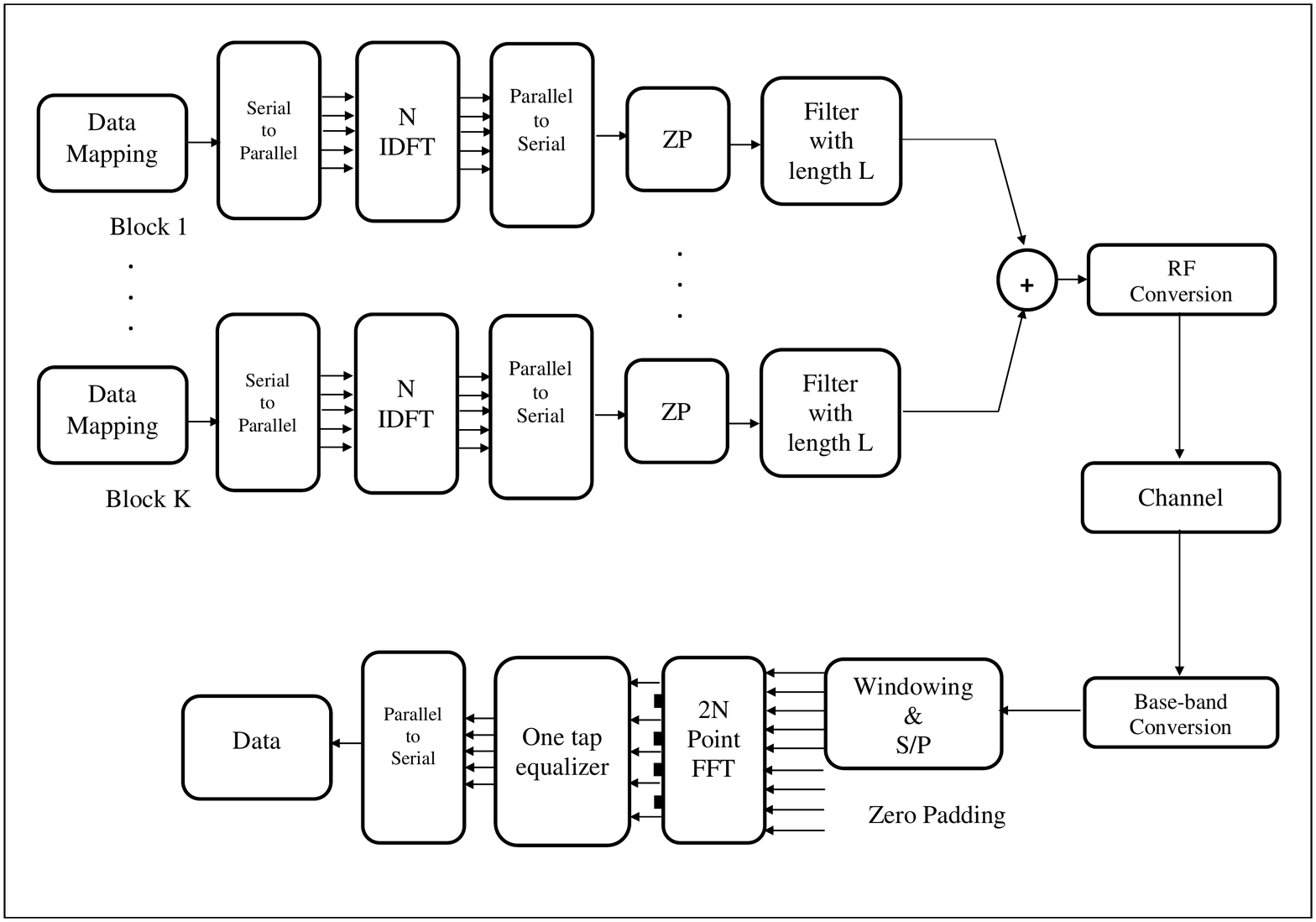}}
\caption{UFMC tranceiver.\label{fig7}}
\end{figure}

\subsection{UFMC}

Universal filtered multi-carrier (UFMC) technique can be considered a derivative of the OFDM with a sub-band filtering operation. This post-filtering process leads to a lower out-of-band leakage than for OFDM. The orthogonality between sub-carriers in each sub-band unit is maintained. Therefore UFMC can do without using offset mapping schemes  like OQAM and by consequence can be used
for MIMO systems. The UFMC transmitter does not involve a cyclic prefix. However, a loss in spectral efficiency is noted due to the operation of linear convolution with the shaping filter (e.g., Dolph-Chebyshev filter)\cite{lynch1997dolph}. The length of the filter is timely short than FBMC one's, improving then the suitability of UFMC for short bursts of communication. \textcolor{blue}{ Fig.\ref{fig7}} shows the block
diagram of an UFMC transceiver. In the $T_x$ stage, the total bandwidth is splited up to several  sub-bands, each one containing a number of sub-carriers  loading by QAM symbols. The result feeds an IDFT of size $N$ followed by the filtering operation. The  UFMC signal is obtained by the superposition of the sub-band-wise filtered components. The  UFMC transmitter performs linear convolution of the IFFT output with a band-pass filter, but this implementation seems to be materially expensive. Another implementation method which reduces substantially the complexity consists of performing the filtering process in frequency domain \cite{gerzaguet20175g}. The receiver stage is composed of a $2N$ point FFT, at the output of this block only even bins are considered to retrieve the data, the odd bins are loaded by inter-carrier interference (ICI) and they are not taken into consideration. Accordingly, a single tap equalizer is sufficient to recover the data. The discrete base-band UFMC signal is mathematically expressed as:

\begin{eqnarray}
x_{UFMC}(n)= \sum\limits_{k=0}^{K-1} S_k(n)*f_k(n)
\end{eqnarray}

Where $K$ is the number of sub-bands, $f_k(n)$ represents the filter coefficients in sub-band of order $k$ and $S_k(n)$ refers to the equivalent OFDM modulated signal over sub-band of order $k$ expressed as:

\begin{eqnarray}
S_k(n)= \sum\limits_{m=0}^{M-1} s_{k,m}(n-m(N+N_g))
\end{eqnarray}

The term $M$ denotes the number of OFDM symbol blocks, and $N_g$ is the length of the zero padding (ZP) added to each symbol.

\begin{eqnarray}
s_{k,m}(n)= \sum\limits_{l=(k-1)L}^{kL-1} A_{l,m,k}\times e^{\mathrm{j \frac{2 \pi ln}{N}}}
\end{eqnarray}

Where $A_{l,m,k}$ denotes the data (e.g: QAM symbols) transmitted over the $l^{th}$ sub-carrier, the $m^{th}$ temporal symbol block and the $k^{th}$ sub-band. A reconfigurable architecture of UFMC with flexible numerology with its FPGA prototype extension is described in \cite{kumar2019reconfigurable}.

\subsection{GFDM}
The General frequency division multiplexing (GFDM)  technique is a block-based techniques where data symbols are spread across two-dimensional structure (time and frequency). It transmits the data symbols during several times slots using different  sub-carriers.  This technique follows the same principle of FBMC by filtering each sub-carrier individually. A prototype filter is translated in time and  frequency to filter each of $N=KM$ QAM complex symbols (one block) to be transmitted over $K$ sub-carriers and $M$ time slots.  So by choosing a proper pulse shape, as root raised cosine (RRC) filter, we can ensure a low OOB emissions. This time-frequency filtering leads to a non-orthogonal waveform, but at the same time allows high-level of flexibility enabling GFDM to be used for sporadic communication \cite{michailow2012generalized}. To avoid the long filter tails of FBMC, GFDM filters each sub-carrier using a circular convolution operation with the RRC filter of length $KM$. The circular convolution process allows keeping the signal length unchanged before and after filtering. This circularity is also known as tail biting technique. A cyclic prefix is inserted in each block enabling GFDM to address the issue of inter-symbol interference.  Thereby, GFDM reaches higher spectral efficiency than OFDM where CP is added to each symbol. The GFDM transceiver architecture is shown in \textcolor{blue}{ Fig.\ref{fig8}}. In the $T_x$ stage, the $M$ QAM symbols to be transmitted in each sub-carrier are up-sampled by factor $K$, after that a convolution is performed with the circular time and frequency shifted version of the prototype filter, the outputs are then modulated with a sub-carrier center frequencies and superposed. This direct implementation seems to be  potentially complex; a wisely reformulating of the GFDM transmitter based on IFFT/FFT approach saves considerably the complexity \cite{farhang2016ofdm}.  The GFDM receiver follows several architectures, namely the zero-forcing (ZF), matched filter (MF), and minimum mean square error (MMSE). The ZF receiver applies the inverse transmitting matrix to the received signal. The transmitting matrix contains operations of up-sampling, pulse shaping, sub-carrier up-conversion and superposition. For the case of the matched filter scheme, after the reception of the whole  block, this one is filtered by the same translated filter used during the $T_x$ stage. The MMSE technique attempts to counteract the noise amplification observed in ZF equalizers by balancing the variance of the noise samples and the data symbols \cite{farhang2015low}. The discrete GFDM signal carried by one block can be expressed as : 

\begin{eqnarray}
x_{GFDM}(n)= \sum\limits_{k=0}^{K-1} \sum\limits_{m=0}^{M-1} d_{k,m}(n)\times g_{k,m}(n),~~ 0\leq n \leq KM-1
\end{eqnarray}

Where $d_{k,m}$ is the data symbol carried by $k^{th}$ 
sub-carrier at  $m^{th}$ symbol and $g_{k,m}(n)$ is the circular shifted version of the prototype filter in time and frequency defined as:  

\begin{eqnarray}
g_{k,m}(n)=  g((n-mK)_{KM}) e^{\frac{\mathrm{i}2\pi kn }{K}},
\end{eqnarray}

Where $g(n)$ is the prototype pulse shaping filter and the term $(.)_{KM}$ is for the modulo operation of coefficient $KM$.

\begin{figure}[!t]
\centerline{\includegraphics[width=250pt,height=20pc]{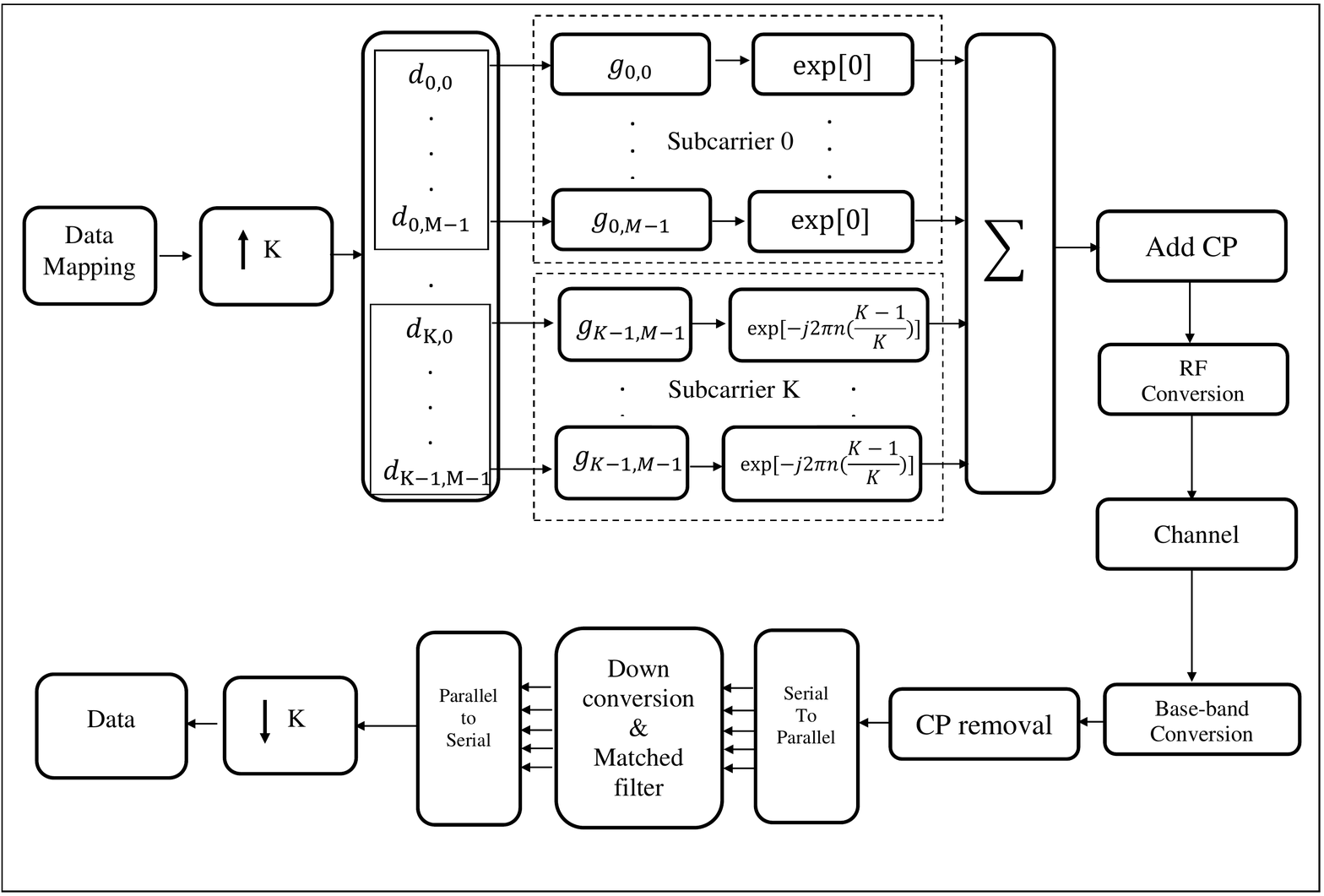}}
\caption{GFDM tranceiver.\label{fig8}}
\end{figure}

\subsection{F-OFDM}

Filtering is considered as a promising solution for reducing side lobes leakages. F-OFDM is one of techniques that introduce filtering operation to overcome the drawbacks of OFDM waveform. The F-OFDM technique is based either  on a whole band filtering or a sub-band based splitting and filtering as with resource block F-OFDM (RB-F-OFDM) \cite{li2013resource}. Two types of filters were considered for generating F-OFDM signal: In the case of the soft-truncated sinc filter, the sinc function is soft-truncated with different window functions, as a result the impulse response vanishes promptly and the ISI is circumscribed \cite{zhang2015filtered}. In the second type, the equiripple Filter structure uses the Remez algorithm associated with equiripple filters to get a sharper transition band so that to alleviate the inter sub-band interference \cite{zhang2015filtered}.

\onecolumn
\begin{center}
\begin{table}
\hspace{20cm}
\centering
\caption{Parameters for the waveforms simulation.\label{tab3}}%
\begin{tabular*}{500pt}{@{\extracolsep\fill}lcclcc@{\extracolsep\fill}}
\hline\\
\textbf{General Parameters}& & &\textbf{OFDM}\\\\
\hline\\
Channel Bandwidth  & & 10MHz &Occupied Subcarriers & & 600\\\\   
Sub carrier spacing & & 15KHz & Guard Subcarriers& & 424\\\\   
Sampling Frequency& & 15.36MHz & Symbol duration & & 66.67 us\\\\  
Number of sub-carriers& & 1024&Number of Resource Blocks& & 50\\\\ 
Cyclic prefix length&& 72 & Number of subcarrier per RB&& 12 \\\\
bits Per SubCarrier && 4 \\\\

\hline\\
\textbf{FBMC} & & &\textbf{UFMC}\\\\
\hline\\
Overlapping factor & & 4 &Guard interval length & &  72\\\\
Prototype filter& & Phydias & Sub-band width & & 12\\\\
Filter length & & 4096 & Chebyshev filter length& & 73\\\\
Number of Resource Blocks& & 50 &Side-lobe attenuation& & 40dB\\\\

\hline\\
\textbf{GFDM} & & &\textbf{F-OFDM}\\\\
\hline\\

GFDM block length & & 15& Filter length & & 513\\\\
Number of guard symbols & & 2& Number of RB& & 50\\\\
Receiver type & & MF & Number of SCs in 1 RBs & & 12\\\\ 
RRC filter roll-off & & 0.2& Tone offset& & 2.5\\\\
Number of IC iterations & &  4& Filter type & & Windowed Sinc (RRC)\\\\

\hline
\end{tabular*}
\end{table}
\end{center}


\begin{center}
\begin{table}
\centering
\caption{Waveforms spectral efficiency.\label{tab4}}%
\begin{tabular*}{500pt}{@{\extracolsep\fill}|l|cccccc@{\extracolsep\fill}}
\hline\\
\textbf{Waveform}& \textbf{Time Overhead}& \textbf{Guard band } & \textbf{Time efficiency } & \textbf{Frequency efficiency }& \textbf{Spectral efficiency }\\
& ($T_{G}$)&($N_{G}$)&\textbf{TE}($M=1$)&\textbf{FE} $(M=1)$&\textbf{SE} $(M=1)$\\
\hline
OFDM& $M \times T_{CP}$& 6M & 0.8928 & 0.9009  & 0.8043 \\\\
FBMC&N $\times (K - 1/2)$ &2M&  0.2808 &0.9966 &0.2799\\\\
UFMC& $M \left( T_{ZP}+(L_{UFMC}-1) \right)$ & 2M & 0.8064 & 0.9966 & 0.8036 \\\\
GFDM & $\frac{M}{N_B} (N_BT_{GS}+T_{CP})$&2M & 0.9937 & 0.9966& 0.9903 \\\\
F-OFDM & $M \times T_{CP}$& 6M & 0.8928 & 0.9009  & 0.8043 &\\

\hline
\end{tabular*}
\end{table}
\end{center}

\twocolumn


 The F-OFDM filter length is relatively short and the sub-carriers are quasi orthogonal by comparison to GFDM waveform. \textcolor{blue}{ Fig.\ref{fig5}} depicts the F-OFDM transceiver. The system is based on an OFDM structure. The transmitter generates the OFDM signal based on the $N$-Point IFFT  and a CP insertion blocks. The F-OFDM signal is then obtained by passing the result through an appropriately designed spectrum shaping filter. At the $R_X$ side, the received signal is first passed through a matched filter block and then the resulting signal feeds a regular OFDM receiver. The F-OFDM signal  can be expressed as:

\begin{eqnarray}
x_{F-OFDM}(t)= x_{OFDM}(t)* F(t)
\end{eqnarray}

Where $F(t)$ is the used shaping filter.

\section{Performance comparison}\label{sec3}

This section is dedicated to the comparison between the filtered waveforms. The comparison is based on several key performance indicators. These are representative for the selection between waveform candidates. The performance of the different waveforms is assessed  in terms of their potential to fill the  major requirements of new communication networks. The simulation parameters are taken under the LTE specifications \cite{sesia2011lte}. These ones are listed in \textcolor{blue}{ Table.\ref{tab3}}. 

\subsection{Power Spectral Density}

The power spectral density (PSD) describes how the power of a signal is distributed in frequency range. It is an important factor that indicates the OOB radiation's level and the ability of a given waveform to reuse the spectrum  and to support coexistence of different services. Therefore, a lower OOB enables users to access network in an asynchronous mode and to incarnate a high mobility.       
The different PSDs are depicted in \textcolor{blue}{ Fig.\ref{fig9}}. The FBMC-OQAM waveform with an overlapping factor $K=4$ represents the best spectral localization. For short burst transmission, a lower pulse duration can be obtained by reducing the overlapping factor $K$ of the prototype filter. However, the spectral containment of the FBMC-OQAM is less  efficient as the $K$ factor decreases. 

Indeed, the OOB of the FBMC-OQAM is $126$dB, $116$dB and $8$dB lower than the OFDM one for an overlapping factor $K=4$, $K=3$ and $K=2$ respectively. The recent work \cite{kobayashi2018fbmc} introduces a new prototype filter design for FBMC-OQAM based on convex optimization. It is capable to supply a high symbol reconstruction with best spectral feature. GFDM has a slightly lower out-of-band leakage compared to OFDM (about $12$dB) but it is clearly outperformed by UFMC.

The OOB for GFDM can be reduced further by using a windowing operation, hereby it approaches UFMC performances. In \cite{de2018linear} the flexibility of the GFDM transmit matrix is examined in order to generate a linear GFDM waveform where good spectral containment of FBMC is achieved. The UFMC technique filters each block separately and  as a result has a lower OOB leakage compared to OFDM, but it is outperformed by FBMC-OQAM where the filtering is applied to each sub-carrier. As for the F-OFDM, there is one filter applied on the whole bands which reduce considerably the OOB. From \textcolor{blue}{ Fig.\ref{fig9}}, one can grasp that OFDM owns the worst worst PSD. To deal with, a spectral encapsulation is introduced in \cite{kim2020spectral} to shape the spectrum of OFDM.

\begin{figure}[t]
\centerline{\includegraphics[width=250pt,height=20pc]{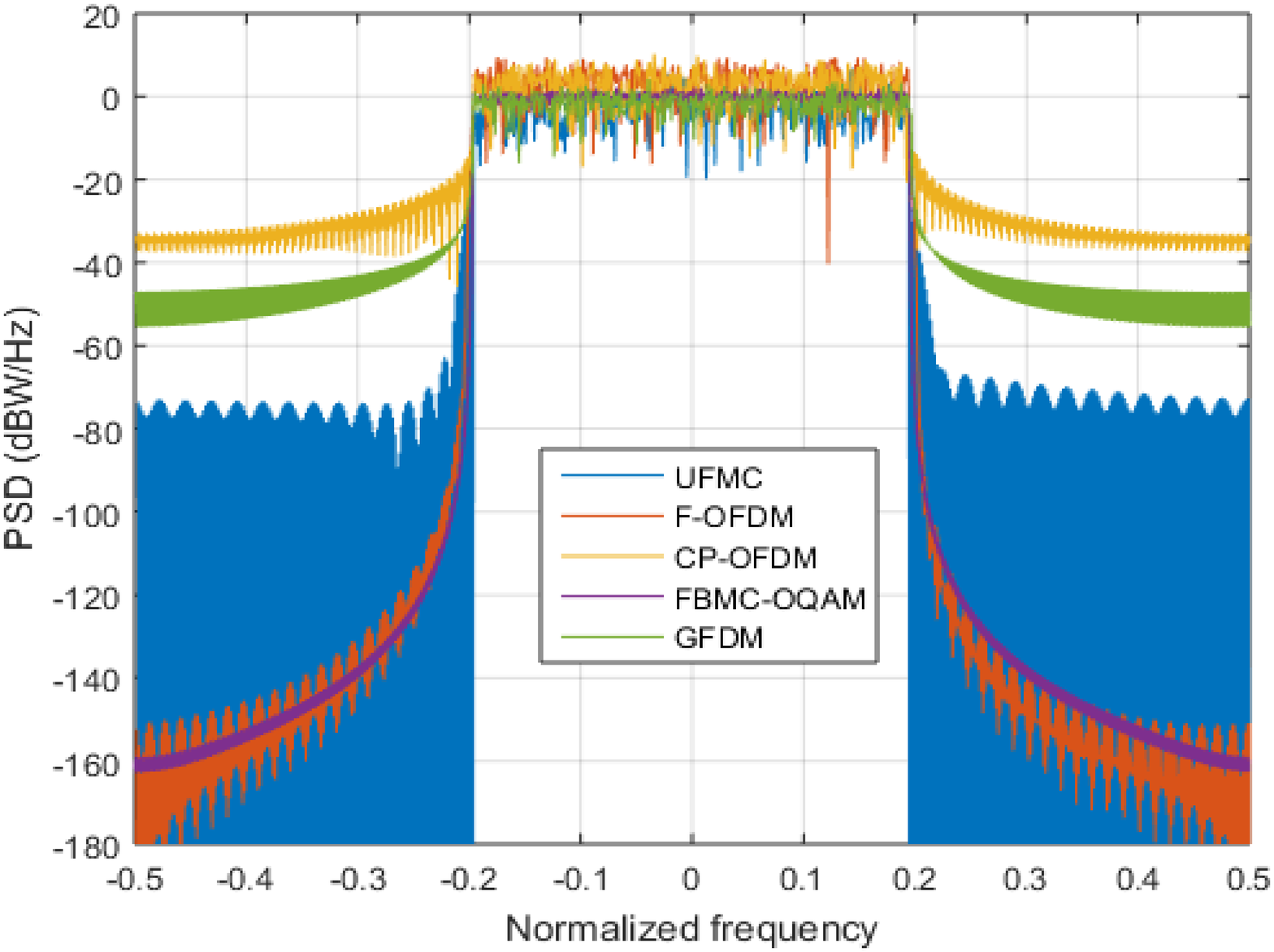}}
\caption{Power spectral density.\label{fig9}}
\end{figure}

\subsection{Spectral Efficiency}

The spectral efficiency parameter represents the number of bits transmitted over a time-frequency unit. This parameter is the product of the time efficiency (TE) and  frequency efficiency (FE). The frequency efficiency characterizes the localization and the spectrum containment of a given waveform in the frequency domain, while the time efficiency quantifies the time overhead introduced to the data symbol.  The frequency efficiency depends on the number of active sub-carriers and the number of guard sub-carriers. The time efficiency is function of the number of samples of symbols in a data burst and the number of overhead samples (CP, filter tail or pilot symbols, etc.). The spectral efficiency represents a key parameter to assess the performance of modulation schemes, it is defined by \cite{schaich2014waveform}:

\begin{eqnarray}
SE= \frac{N_{D}}{N_{D}+ N_{G}} \times \frac{T_{D}}{T_{D}+ T_{G}}
\end{eqnarray}

Where $N_D$ and $N_G$ are respectively the number of data sub-carriers ($600$ in LTE) and guard band sub-carriers, on the other hand, $T_D$ and $T_G$ represent the data and time overhead samples. The TE, FE and related SE for each waveform are summarized in \textcolor{blue}{ Table.\ref{tab4}} for $M$ multi-carrier symbols. The CP inserted in OFDM and F-OFDM  constitutes the only overhead samples. The Time-domain overhead for OFDM and F-OFDM is proportional to CP and burst length, thus their time efficiency is depending only on CP length. The FBMC system incurs an overhead due to the filter tails and time offset between the OQAM symbols. It uses inefficiency spectral resources because of these long ramp-up and
ramp-down tails especially for short bursts. A novel method for shortening these tails is set forth in \cite{qu2017improving} and therefore improving spectral efficiency. The time efficiency in FBMC-OQAM is independent from the frame size. GFDM represents the highest efficiency due to its block-based feature using a unique CP per frame. Otherwise, UFMC introduces a filter tail in each block and has an overhead equals to the summation of the filter length and the zero padding samples.  According to the LTE standard, $666$ sub-carriers of $15$KHz wide fit the $10$MHz bandwidth and the number of sub-carriers carrying data is $600$. The FBMC-OQAM,GFDM and UFMC are designed to achieve a narrow OOB radiation and thus require few guard bands (one on each side). OFDM and F-OFDM do not exhibit very low OOB, in these two cases, the guard band is expected to be more large. \textcolor{blue}{ Fig.\ref{fig10}} evinces that all waveforms exhibit a constant spectral efficiency versus the number of multi-carrier symbols except the FBMC-OQAM, this one depends on the frame duration and its performance approaches that of the UFMC as the number of symbols per frame increases. GFDM is the most spectral efficiency waveform thanks to it well frequency localization and thanks to the tail biting technique. Due to the constant filter tails, UFMC shows a time efficiency that approaches that of the OFDM with a slight improvement due to lower guard band required for UFMC. F-OFDM and OFDM show the same performances since using hard truncation.

\begin{figure}[h]
\centerline{\includegraphics[width=250pt,height=20pc]{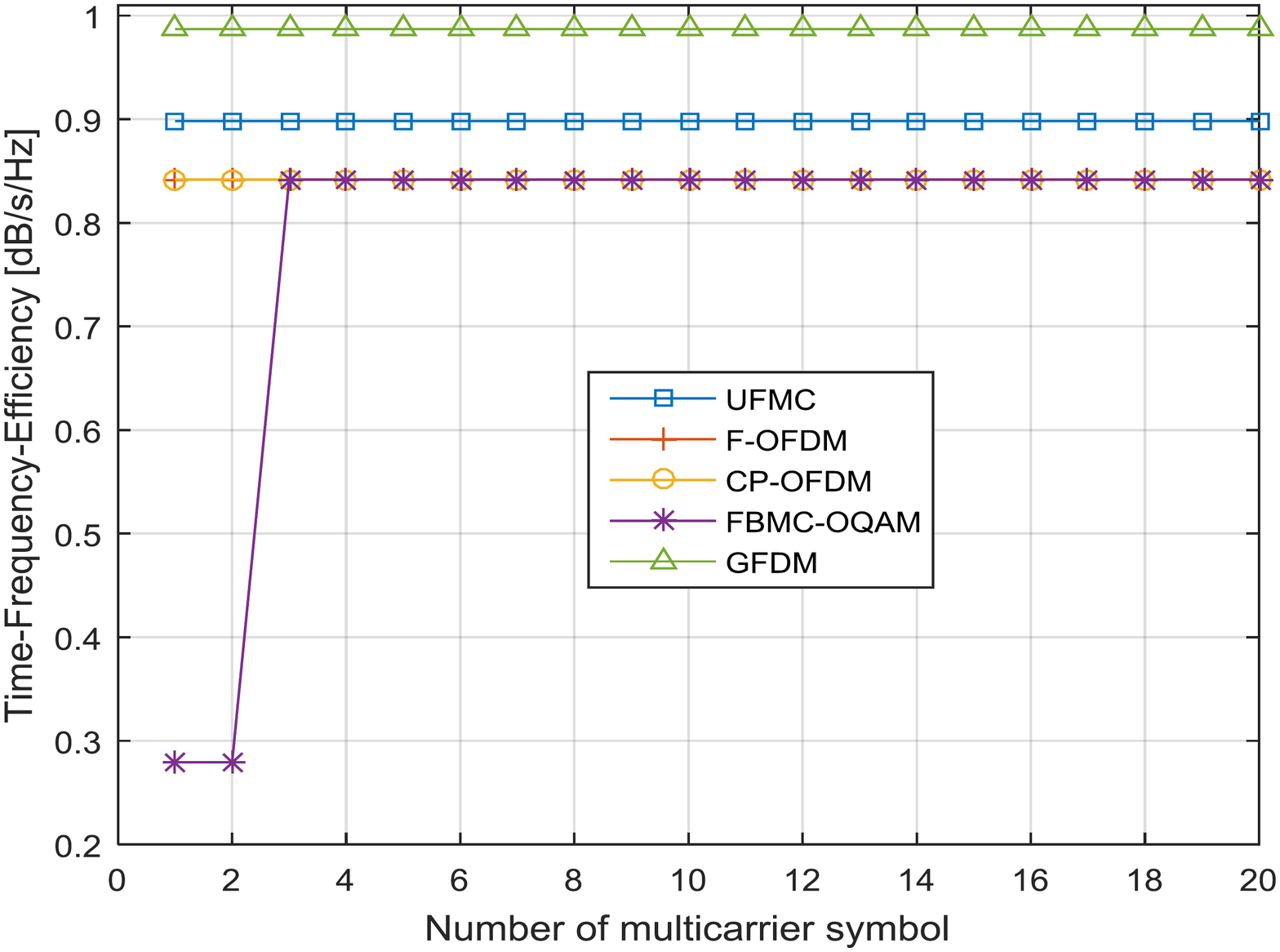}}
\caption{Spectral efficiency.\label{fig10}}
\end{figure}

\subsection{Bit Error Rate  }

The performance of the considered waveforms in terms of bit error rate (BER) is assessed using the Extended Vehicular A channel model (EVA) defining a medium  delay spread environment. The EVA channel model represents a multi-path delay profile used to measure performance in multi-path fading environment. A multi-path fading propagation conditions are defined by a combination of a multi-path delay profile (EVA) and a large Doppler frequency.  Here we set the Doppler frequency to $5$ Hz. \textcolor{blue}{ Fig.\ref{fig11}} shows the BER versus SNR for the EVA channel. We assume for the simulation that the channel state information (CSI) is perfectly known and no  synchronization misalignment is considered. The receiver uses a hard decision. Thanks to the samples of the CP guard interval , the CP-OFDM and F-OFDM implementations show an excellent performance. The FBMC-OQAM due to the long filter length outperform GFDM performance. Decreasing the size of FFT block in the FBMC-OQAM implementation degrades the performance in terms of BER, this is mainly caused by the resulting reducing in the filter length in the time domain. In this moderately frequency selective channel, CP-OFDM and UFMC do likewise. GFDM implementation exhibits the worse BER compared to other waveforms. The BER level seems to reach high-level with high constellation order for all considered waveforms. In\cite{mohanraj2019performance} local discrete Gabor transform (LDGT) algorithm is used with a new type of prototype filter to eliminate interference in GFDM system. The proposed filtering introduces the orthogonality condition and have separate filters for the even and odd subcarriers respectively. 

\begin{figure}[b]
\centerline{\includegraphics[width=250pt,height=20pc]{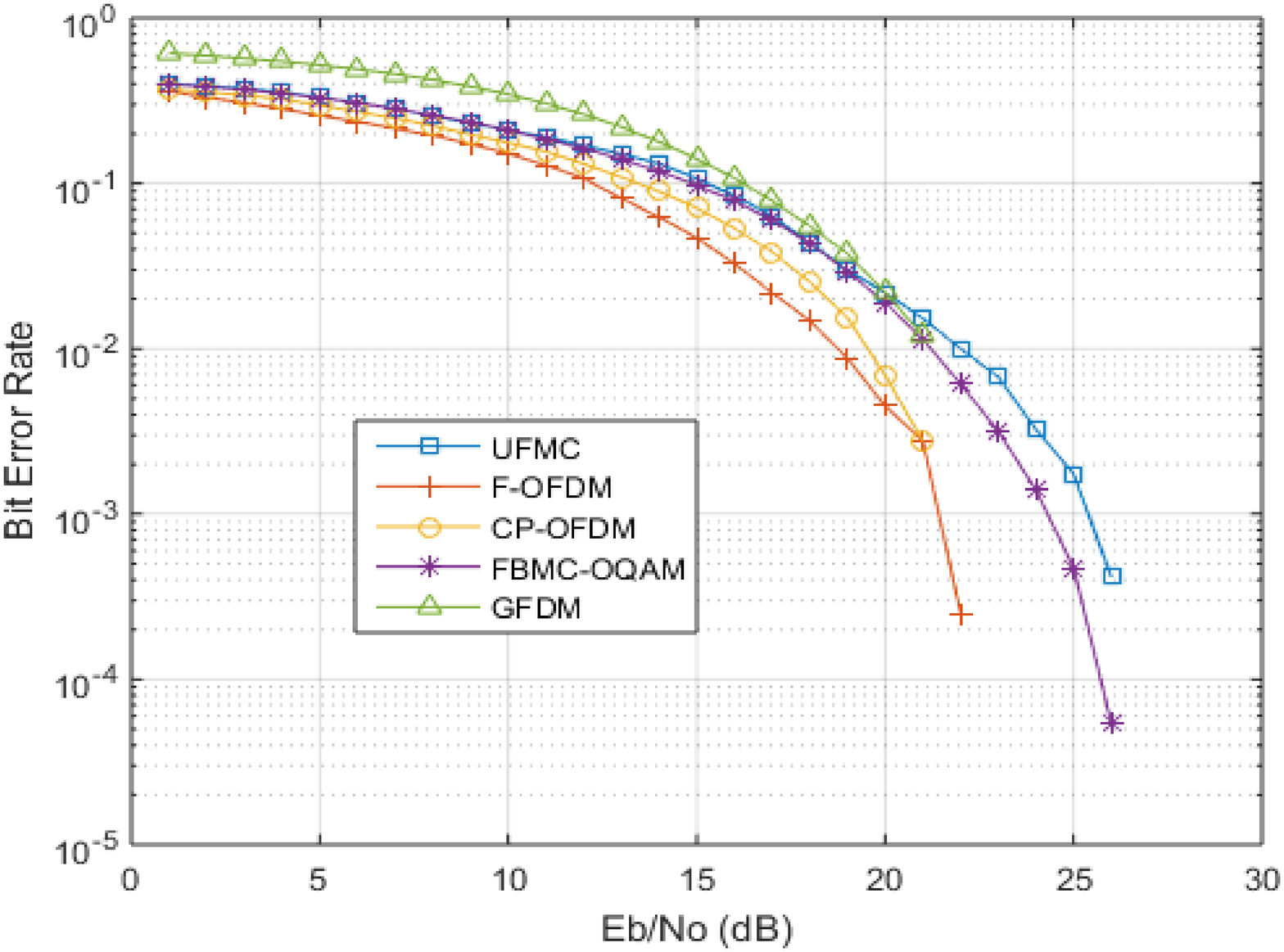}}
\caption{Bit error rate.\label{fig11}}
\end{figure}

\subsection{Energy Efficiency}
 Information and communication technology industries are contributing more and more to the CO$_2$ emission because of their high energy consumption. This energy is used to serve the great demand on electricity, especially by the base station (BS) \cite{bogucka2011degrees}. Therefore, one of the main challenges of 5G and B5G communication systems is to enhance their energy efficiency. The peak-to-average power ratio (PAPR) is a critical parameter for describing waveforms energy efficiency.  Waveforms having high PAPR limit the Power amplifier (PA) to operate in its linear region. This    restraint reduces significantly Power amplifier efficiency (PAE). Contrarily, waveforms with low PAPR allow PA to operate near saturation region, reaching thus high efficiency. PAPR represents the ratio of the peak power of the signal to its average power. The PAPR is defined as :
 
 \begin{eqnarray}
PAPR= \frac{Max | x(t)|^2}{E \left[ | x(t)|^2 \right] }
\end{eqnarray}

Since PAPR is viewed as a random variable, the PAPR description of the filtered waveforms is provided by the complementary cumulative distribution function (CCDF). This one is defined for a threshold $\gamma$ by:

 \begin{eqnarray}
CCDF_{PAPR}= Pr_{\gamma} \left(PAPR>\gamma \right)
\end{eqnarray}

 We compute on \textcolor{blue}{ Fig.\ref{fig12}}, the CCDF of the PAPR for each waveform. The filtered waveforms show all a high PAPR, this is due to their multi-carrier nature. The OFDM waveform detains the best performance, while F-OFDM shows the highest PAPR about $1$dB  compared to OFDM. The Filtering operation used in F-OFDM is the major contributors to PAPR increase. FBMC-OQAM, GFDM and UFMC even they show a high PAPR than OFDM ($0.4$dB), they remain slightely better than F-OFDM. Several PAPR reduction techniques have been proposed to deal with the high PAPR of multi-carrier systems. Distorting the signal before the amplification processing can reduce PAPR, clipping and filtering \cite{panta2004effects}, peak windowing \cite{kim2005new} and companding \cite{aburakhia2009linear} belong to the signal distortion techniques class. Other techniques generate multiple permutations of the multi-carrier signal. They transmit the one with minimum PAPR or modify the  signal parameters like the phase or constellation to achieve a low PAPR. These techniques like Tone reservation (TR)\cite{wattanasuwakull2005papr}, Tone injection \cite{wattanasuwakull2005papr} and Partial Transmit Sequence (PTS) \cite{muller1997ofdm} are called  Probabilistic Techniques. There is a third class based on coding the signal sequences, this one is called coding techniques class. The all mentioned techniques are well established for OFDM systems, but not yet completely investigated for the other considered waveforms. 
 
\begin{figure}[b]
\centerline{\includegraphics[width=250pt,height=21pc]{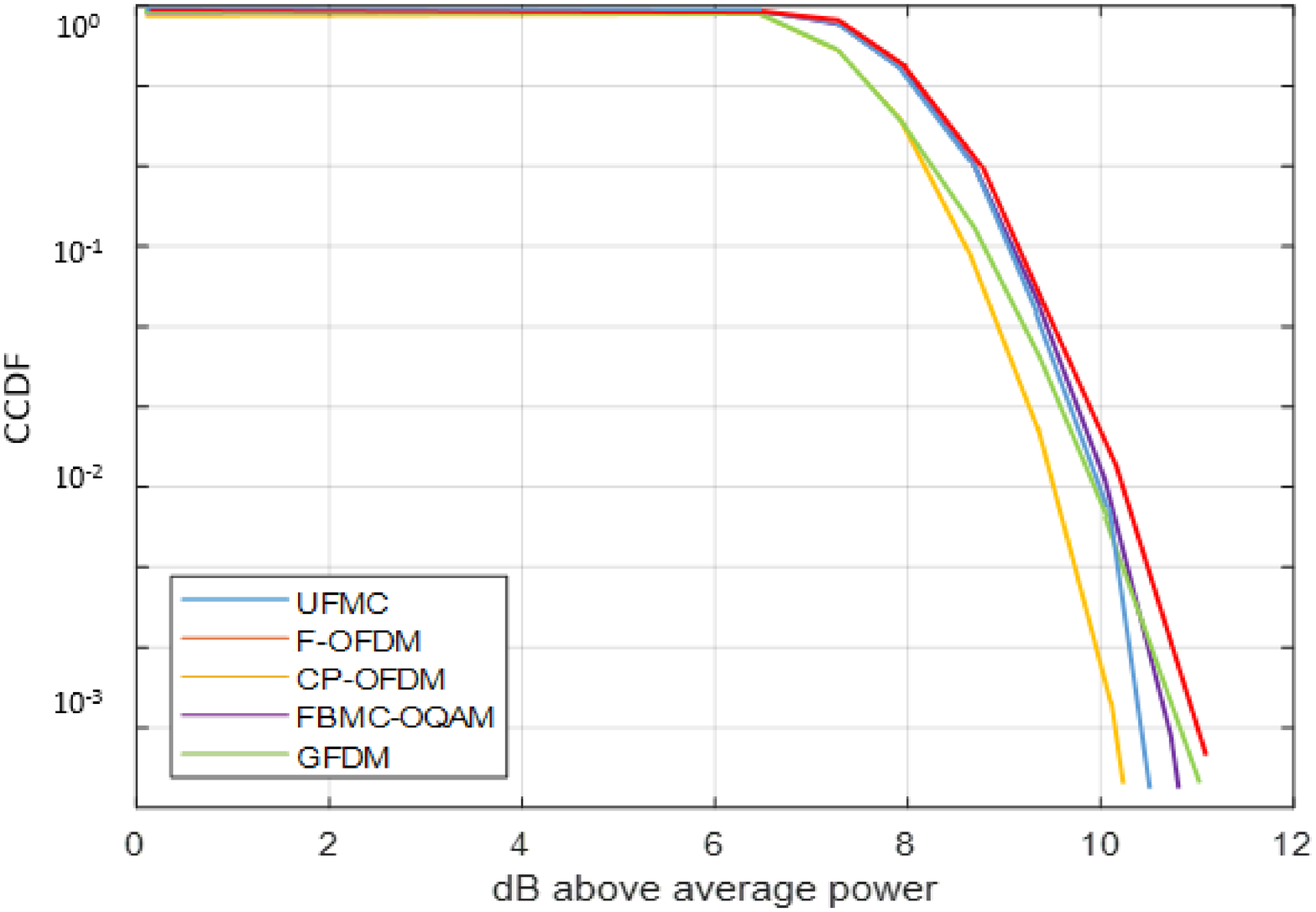}}
\caption{Complementary cumulative distribution function.\label{fig12}}
\end{figure}

\subsection{Complexity}

The computational complexity for a given waveform is a key parameter to assess how they can be executed by processors. The presented complexity is only a fraction of the overall system complexity, RF digital pre-processing and encoders consume a lot of operations as well. The complexity is  evaluated in terms of the number of real multiplications per multi-carrier symbol. This one is used  as a metric for modulator complexity. The transceiver complexity of the different waveforms are summarized in \textcolor{blue}{ Table.\ref{tab5}}. The values in \textcolor{blue}{ Table.\ref{tab5}} don't take the complexity of the equalizer into account. For CP-OFDM system, the required complexity  to generate the signal is reduced to the IFFT block operation at the transmitter and to an FFT block operation at the receiver. If we do not take in consideration the complexity of the equalizer, the complexity of OFDM transmitter/receiver only equals the number of real multiplications in IFFT/FFT stages. Thus, the OFDM transceiver complexity is the double of one stage. OFDM represents the lowest computational complexity between all waveforms. The split-radix algorithm is considered for the IFFT and FFT implementation. The FBMC technique controls  the frequency response by introducing a filter bank centered on the active sub-carriers and based on the same prototype filter. For the poly-phase decomposition structure, an efficient implementation is obtained by deploying an offset-QAM staggering, a phase rotation for linear phase, an IFFT/FFT and a poly-phase filtering \cite{baltar2011computational}. The FFT block in transmitter side is fed with  pure real or imaginary symbols (OQAM symbols) which result in less complexity. However, the receiver FFT block uses complex inputs/outputs symbols. FBMC technique is nearly six times more complicated than OFDM. A straightforward implementation of the UFMC technique in frequency domain consists of summing up all sub-bands with per sub-band operations of IFFT and filtering, but this direct  implementation is not efficient.  An efficient implementation method of the UFMC modulator is proposed in \cite{wild2015reduced} using overlapping sub-bands. Each sub-band is proceeded by small size IFFT/FFT, the filtering operation is performed in the frequency domain and the overlapping sub-bands are superimposed in a large $2N$-IFFT block, in which the number $2$ represents the over-sampling factor. Each sub-band requires an IFFT of size $N$ and an FFT of size $2N$. The frequency filtering operation requires $8N$ real multiplications. The $R_x$ stage includes a windowing in the time domain, FFT block of size $2N$ with zero padding and a frequency domain filtering. With this structure, UFMC generates the highest complexity compared to the other waveforms. The GFDM technique is based on a two-dimensional data structure, data symbols are grouped in blocks containing each one $N$ sub-carriers and $M$ time slots. The processing of these blocks is  based on digital filters with tail-biting property that preserves circular properties across frequency and time domain. The generation of GFDM signal requires a domain conversion in which data signal of each sub-carrier is converted into the frequency domain. Then, after an up-sampling operation, a filtering process is applied in the frequency domain followed by a frequency up-conversion. Finally, the signal is re-transformed  into the time domain \cite{gaspar2013low}.  The receiver contains transformation of the signal into the frequency domain, a frequency  filtering and a re-transformation to time domain \cite{gaspar2013low}. A recent low complexity modem design for GFDM waveform is proposed in \cite{farhang2015low}. Due to the circular filtering, GFDM is far less complex than UFMC. The Filtered OFDM aims to improve the spectral containment of the traditional CP-OFDM using a filter at the output of a CP-OFDM transmitter. In addition to the IFFT block, the complexity of F-OFDM system entails more terms induced by the filtering operation. The used filter is real and the filtering operation is followed by an up-conversion that contribute to the complexity with an extra real multiplication \cite{medjahdi2017road}. In \cite{kim2019low}, singular value decomposition based F-OFDM (SF-OFDM) system is proposed with much short filter length than F-OFDM system. This approach reduces considerably the burden of the F-OFDM system.  

\subsection{Resilience to PA non-linearity}

Power amplifiers (PA) play a key role in both transmitter and receiver sides of a communication system. They bring the desired signal to a suitable level of transmission or for demodulation. PAs introduce two main impairments due to their non ideal nature: nonlinearity and additive noise. The power amplifier  is the most frequent source of non-linearity in the transmitter. The nonlinear behavior of an amplifier can be described in a general manner by a power series expansion  \cite{horlin2008digital}.  The nonlinear feature of PAs introduces both an out-of-band spectral regrowth and in-band distortion. The in-band interference is usually characterized by the error vector magnitude (EVM).
 The spectral regrowth comes especially from the convolution operation involved in the Fourier transform expression of the non-linearity model. The robustness to PA non-linear distortions is assessed using the output back-off (OBO) parameter. To check the resilience of filtered waveforms to PA non-linearity, we use the Saleh model and the modified Rapp model.  The Saleh model has the following amplitude (AM-AM) and phase (AM-PM) characteristics \cite{gharaibeh2011nonlinear}:

 \begin{eqnarray}
f_{S_{AM/AM}}(x)= \frac{A.x}{ 1+B. x^2 } 
\end{eqnarray}

\begin{eqnarray}
f_{S_{AM/PM}}(x)= \frac{C.x^2}{1+D.x^2} 
\end{eqnarray}

Where $x$ denotes the amplitude of the input signal and $A, B, C, D$ are PA parameters. The Rapp model is based on experimental measurements with an AM-AM and AM-PM characteristics given by \cite{gharaibeh2011nonlinear}:

 \begin{eqnarray}
f_{R_{AM/AM}}(x)= \frac{G.x}{ \left( 1+ \left( \frac{G.x} {V_{SAT}} \right)^{2P} \right)^{ \frac{1}{2P}}} 
\end{eqnarray}

\begin{eqnarray}
f_{R_{AM/PM}}(x)= 1
\end{eqnarray}

 The AM/AM and AM/PM characteristics describe how the instantaneous input amplitude influences the output amplitude and phase, respectively. The spectral regrowth of the different filtered waveforms due to Power amplification operation are depicted in \textcolor{blue}{ Fig.\ref{fig13}} for the modified Rapp models. It is clear that the type of the model affects the OOB spectral regrowth level of the waveform passing through the PA. This out-of-band distortion can significantly reduce the system performances in a multi-users access configuration.  The out-of-band distortions are investigated by calculating the adjacent channel power ratio (ACPR). The results are shown in \textcolor{blue}{ Fig.\ref{fig14}}. The in-band distortion are measured using the error vector magnitude. The EVM is defined as the normalized magnitude of the difference between the input and output of a non-linearity. It  measures the extent of the departure of the signal constellation from the ideal reference. The constellation mapping or BER are also used as in-band distortion indicators (\textcolor{blue}{ Fig.\ref{fig11}}). 

\begin{figure}[t]
\centerline{\includegraphics[width=250pt,height=20pc]{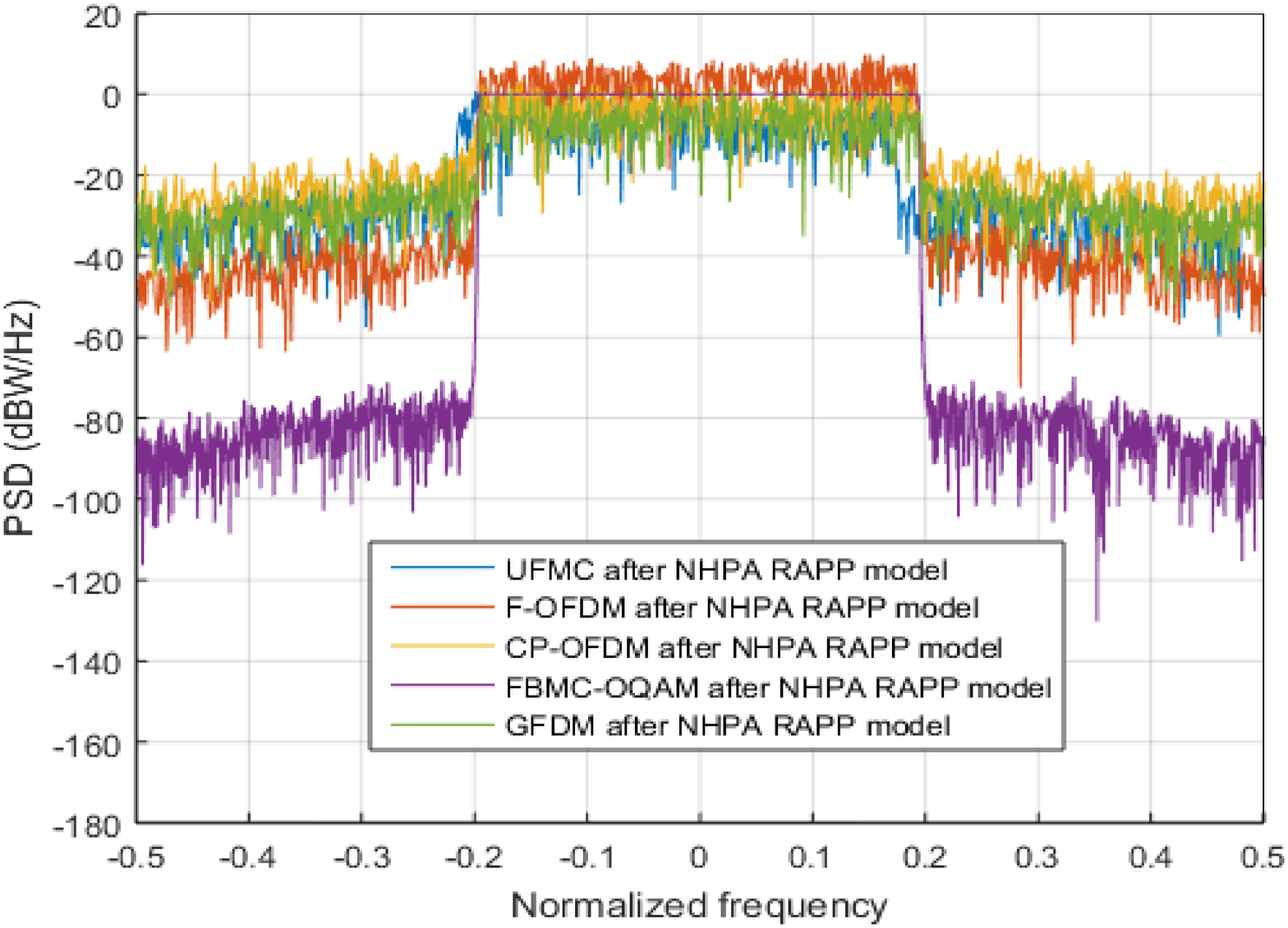}}
\caption{Nonlinear Power amplifier effect.\label{fig13}}
\end{figure}

\section{Filter-Waveform dependency}
 The present review  deduces the strong relationship between the prototype filter properties and the corresponding filtered waveform performances. The filtering operation, as first goal, tends to enhance the spectrum containment. This alleviates the abundant use of the spectrum and  allows coexistence between systems by enhancing PSDs and reducing OOB leakage.

\onecolumn
\begin{center}
\begin{table}
\centering
\caption{Waveforms Complexity.\label{tab5}}%
\begin{tabular*}{500pt}{@{\extracolsep\fill}lcr@{\extracolsep\fill}}
\hline\\
\textbf{$T_x$ Complexity} & \textbf{Waveform} &\textbf{$R_x$ Complexity}\\
\hline\\
& CP-OFDM   \\\\   
$C_{IFFT}=N \log_2N-3N+4 $&&$C_{FFT}=N \log_2N-3N+4 $\\\\
\hline\\
& FBMC-OQAM (PPN)  \\\\ 
$C_{SFB} = N(\log_2(\frac{N}{2})-3)+8+4(NK + 1)$&& $C_{AFB} = 2N(log_2(N)-3) + 8 + 4(NK + 1)$ \\\\
\hline\\
&UFMC      \\\\  
$ B\left[ C_{IFFT}(N)+ C_{FFT}(2N)+8N \right] + C_{FFT}(2N)$ && $ C_{FFT}(2N)+8N $ 
\\\\
\hline\\
&GFDM       \\\\ 
$ NM \log_2(M)+NLM+ MN \log_2(MN)  $ &&$NM\log_2(NM)+NLM+NM \log_2(M)  $    \\\\
\hline\\
& F-OFDM   \\
$2N \log_2N-6N+8  +2(N+T_{CP})L_{F}+2NL_{F}$ && $2N \log_2N-6N+8  +2(N+T_{CP})L_{F}+2NL_{F}$ \\\\

\hline
\end{tabular*}
\end{table}
\end{center}


\begin{center}
\begin{table}%
\centering
\caption{Papers dealing with filtered waveforms.\label{tab6}}%
\begin{tabular*}{550pt}{@{\extracolsep\fill}|l|cccccc@{\extracolsep\fill}}
\hline\\
\textbf{Waveform}& \textbf{OFDM}& \textbf{FBMC} & \textbf{UFMC} & \textbf{GFDM}& \textbf{F-OFDM}\\\\
\hline\\\\

Channel Estimation & \cite{edfors1998ofdm},\cite{minn2000investigation}& \cite{cui2015coded},\cite{fuhrwerk2017scattered} & \cite{wang2015pilot},\cite{wang2017compressive} & \cite{vilaipornsawai2014scattered},\cite{na2018turbo}  & \cite{zhang2017filtered}\\

& \cite{ma2007scattered},\cite{liu2014channel} &   &   & \cite{jeong2018eigendecomposition} &  \\\\\\\\
Equalization &  \cite{jarajreh2014artificial},\cite{primo2012equalization} &\cite{ihalainen2011channel},\cite{ndo2012fbmc} &  \cite{zhang2016mu},\cite{rani2016ufmc} & \cite{zhong2018iterative},\cite{carrick2017improved} & \cite{zillmann2007turbo}\\
&\cite{panayirci2010joint},\cite{al2012low}& & & \cite{tiwari2017low}
\\\\\\\\
Synchronization& \cite{park2004blind}, \cite{wang2003ser} & \cite{stitz2010pilot}, \cite{medjahdi2011performance}  & \cite{schaich2014relaxed}, \cite{cho2017asynchronous}  & \cite{gaspar2014synchronization}, \cite{wang2016maximum}  &   \\
& \cite{minn2003robust} &   &  &  \cite{na2018pseudo} &  \\\\\\\\
PAPR & \cite{jiang2008overview},\cite{irukulapati2009slm} & \cite{kollar2012clipping},\cite{na2017low} & \cite{rong2017low},\cite{baki2019novel} & \cite{sharifian2015polynomial},\cite{sharifian2016linear} & \cite{mabrouk2017precoding} \\
& \cite{kollar2012clipping}, \cite{rahmatallah2013peak} & \cite{wang2016hybrid} &  & \cite{bandari2017papr} & \\\\\\\\
MIMO compatibility & \cite{pande2007reduced},\cite{mody2001synchronization} & \cite{zakaria2012novel}, \cite{bendimerad2019low}  & \cite{buzzi2019mimo}, \cite{chen2017mimo}  & \cite{matthe2016sphere}, \cite{matthe2017low} &\\
&\cite{yang2005road}& & & \cite{zhang2015expectation} \\\\\\\\
Resource Allocation&  \cite{attar2008interference}, \cite{adian2013optimal} & \cite{zhang2011noncooperative}, \cite{denis2016energy}  & \cite{kim2016resource}, \cite{del2016resource}  &  \cite{mokdad2016radio}, \cite{gonzalez2017resource}  &  \cite{li2014resource}  &\\
& \cite{cai2016subcarrier}\\
\hline
\end{tabular*}
\end{table}
\end{center}


\twocolumn

\begin{figure}[t]
\centerline{\includegraphics[width=250pt,height=20pc]{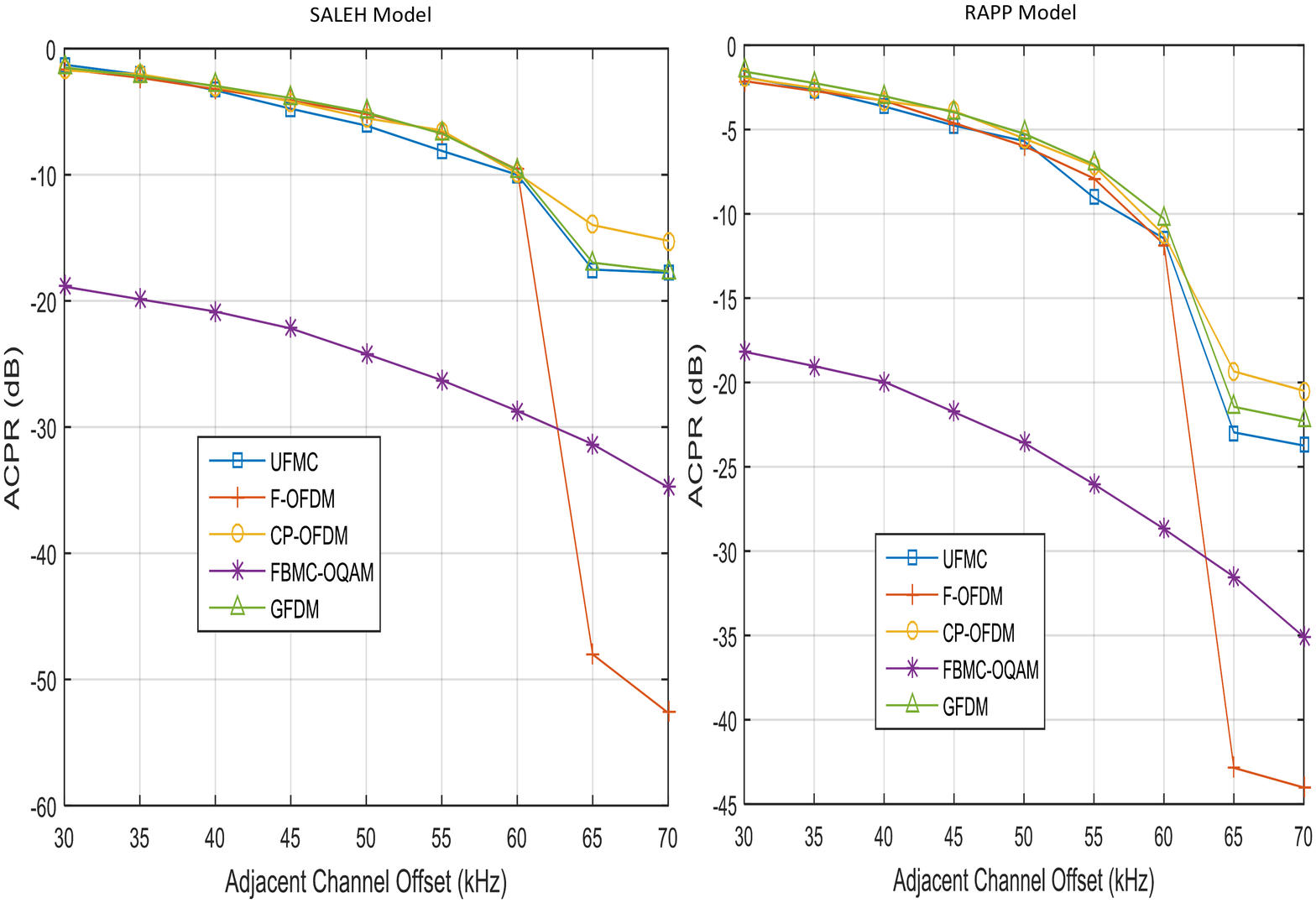}}
\caption{Adjacent channel power ratio.\label{fig14}}
\end{figure}

 The filter decaying property and ambiguity function are  in control of these  characteristics.  Additionally, filtering operation augments the frequency localization and reduces ICI and ISI. The complexity of a given waveform depends directly on the length of the filter and the number of filter taps. The time domaine implementation is generally more complex than in frequency domaine. The waveform latency systematically depends on the length of the filter. Good latency can be achieved through a good time localization of the filter. However, in some cases, the need for a long filter is to avoid ISI and ICI. The circular convolution and oversampling process also alter the latency performance \cite{hammoodi2019green}. The prototype filter affects the spectral efficiency performance as well. A well time-frequency localized pulse shape filter makes the waveform more suitable for a good spectral efficiency. However the length of the filter and its stopband attenuation can result in a reduction of the spectral efficiency. As for PAPR, the prototype filter length has a significant influence on the probability distribution of the PAPR and on the PAPR reduction technique. For example the PAPR increases when the symbols overlap becomes significant due to long prototype filter in the case of FBMC. So, according to this survey, a key element to be undertaken in the design of the 5G and B5G systems is the close relationship between the prototype filters and the performances of the resulting filtered waveforms. These performance parameters can be formulated as typical filter design problematic or as a filter coefficient optimization problem as in \cite{zhao2020genetic}.

\section{Conclusion}\label{sec2}


Due to the diverse applications of new communication networks, the architectural adjustments associated with the radio access layer seem to be mandatory. These ones extend from including new air-interfaces to deployment of new multiple access techniques for heterogeneous networks. The design of prototype filters constitutes an arresting step for the conception of  new air-interfaces like filtered and bank filtered ones. These modulations schemes are based on a number of contiguous and frequency translated bands conceived from the same prototype filter. The ambiguity function represents an important factor used to compare and to differentiate between prototype filter's efficiency in terms of ISI and ICI free transmission.  Filters covered here are designed to satisfy a good time/frequency localization and a near perfect orthogonality. Energy concentration and transition band decay are other important selecting criteria that indicated how the energy is distributed and how fast is the filter transition band slope. These factors are important in determining the spectral regrowth in presence of radio front-end impairments and systems coexistence leakage. The review found that the prototype filter properties (length, transition band,...) affect the performance of the filtered waveforms. The study approves its usefulness in determining the overall complexity, latency, spectral efficiency and BER.  Although most filters show good performances, the choice depends on practical concerns as implementation and numerical complexity. 
Also, new air-interface waveforms have been extensively analyzed and evaluated based on several key performance indicators. Power spectral density, spectral efficiency, BER, PAPR and computational complexity and resistance to PA impairment have been assessed. It is clear from the comparison that all considered waveforms drastically improve the PSD with respect to OFDM. Since all waveforms are multi-carrier schemes, they all have the same PAPR issue. Recently Energy Harvesting and artificial intelligence-based solutions have been proposed to alleviate this problem. The F-OFDM and UFMC are interesting options to replace OFDM as they enhance performance in terms of PSD and of asynchronous access. Furthermore, these waveforms are compatible to almost OFDM algorithms (channel estimation, equalization and spatial diversity). The energy efficiency and weak resilience to PA impairment stay the main drawbacks for UFMC and F-OFDM waveforms. Presenting the best time-frequency containment entitling the use of fragmented spectrum and energy efficiency comparable to that of OFDM ; GFDM and FBMC-OQAM are also high attractive candidates. Even if the FBMC-OQAM exhibits good performance in non-synchronous access, it is not adapted to short packet size. GFDM implementation represent the best performance in terms of SE. However, due to the lack of orthogonality, this waveform had difficulty finding MIMO compatibility. MIMO and massive MIMO ability is an open problem for multicarrier non-orthogonal waveforms. Still, there is a tremendous amount of  research in  this  field  and  early results show that all candidate waveforms have a potential  to  be  implemented  with  MIMO  systems.  The multi-numerology concept, where multiple frame structures with different sub-carrier spacing coexist, is a new reseach direction in this field. Various critical issues concerning multi-numerology system have emerged such as: inter-numerology interference and orthogonality between sub-carriers of different numerologies. GFDM and FBMC-OQAM incarnate a high complexity. This one is still a high topic for these two waveforms. Finally, it seems that there is no definite preferable modulation format. Even if OFDM has been nominated for 5G standard because of its implementation simplicity, the choice will depend essentially on the considered scenarios and the selected parameters, leaving the field open for future proposals and contributions.

\bibliography{cas-refs}%

\bibliographystyle{unsrt}

\nocite{*}



%





\end{document}